\documentclass[onecolumn,11pt]{IEEEtran}
\usepackage{amsmath}
\usepackage{amssymb}
\usepackage{amsfonts}

\newcommand{\Rmnum}[1]{\expandafter\@slowromancap\romannumeral #1@}
\usepackage{graphicx}
\usepackage{epsfig}
\usepackage{subfigure}
\usepackage{citesort}
\usepackage{xcolor}
\usepackage{enumerate}
\usepackage{extarrows}
\usepackage{algorithmic}
\begin{document}
\vspace{-.5in}
\title{Wireless Powered Cooperative Jamming for Secrecy Multi-AF Relaying Networks}

\author{Hong Xing, Kai-Kit Wong, Arumugam Nallanathan, and Rui Zhang
\thanks{Part of this paper has been presented at the IEEE International Conference on Communications (ICC), London, UK, June 8-12, 2015.}
\thanks{H. Xing and A. Nallanathan are with the Centre for Telecommunications Research, King's College London (e-mails: $\rm hong.xing@kcl.ac.uk$; $\rm arumugam.nallanathan@kcl.ac.uk$).}
\thanks{K.-K. Wong is with the Department of Electronic and Electrical Engineering, University College London (e-mail: $\rm kai$-$\rm kit.wong@ucl.ac.uk$).}
\thanks{R. Zhang is with the Department of Electrical and Computer Engineering, National University of Singapore (e-mail:$\rm elezhang@nus.edu.sg$). He is also with the Institute for Infocomm Research, A*STAR, Singapore.}}
\maketitle

\thispagestyle{empty} \vspace{-0.3in}

\begin{abstract}
This paper studies secrecy transmission with the aid of a group of wireless energy harvesting (WEH)-enabled amplify-and-forward (AF) relays performing cooperative jamming (CJ) and relaying. The source node in the network does {\em simultaneous wireless information and power transfer (SWIPT)} with each relay employing a power splitting (PS) receiver in the first  phase; each relay further divides its harvested power for forwarding the received signal and generating artificial noise (AN) for jamming the eavesdroppers in the second transmission phase. In the centralized case with global channel state information (CSI), we provide closed-form expressions for the optimal and/or suboptimal AF-relay beamforming vectors to maximize the achievable secrecy rate subject to individual power constraints of the relays, using the technique of semidefinite relaxation (SDR), which is proved to be tight. A fully {\em distributed} algorithm utilizing only local CSI at each relay is also proposed as a performance benchmark. Simulation results validate the effectiveness of the proposed multi-AF relaying with CJ over other suboptimal designs.
\end{abstract}


\IEEEpeerreviewmaketitle
\setlength{\baselineskip}{1.3\baselineskip}
\newtheorem{definition}{\underline{Definition}}[section]
\newtheorem{fact}{Fact}
\newtheorem{assumption}{Assumption}
\newtheorem{theorem}{\underline{Theorem}}[section]
\newtheorem{lemma}{\underline{Lemma}}[section]
\newtheorem{proposition}{\underline{Proposition}}[section]
\newtheorem{corollary}[proposition]{\underline{Corollary}}
\newtheorem{example}{\underline{Example}}[section]
\newtheorem{remark}{\underline{Remark}}[section]
\newtheorem{algorithm}{\underline{Algorithm}}[section]
\newcommand{\mv}[1]{\mbox{\boldmath{$ #1 $}}}

\section{Introduction}
 Wireless powered communication network has arisen as a new system with stable and self-sustainable power supplies in shaping future-generation wireless communications \cite{Lu2015,Bi2015}. The enabling technology, known as simultaneous wireless information and power transfer (SWIPT), has particularly drawn an upsurge of interests owing to the far-field electromagnetic power carried by radio-frequency (RF) signals that affluently exist in wireless communications. With the transmit power, waveforms, and dimensions of resources, etc., being all fully controllable, SWIPT promises to prolong the lifetime of wireless devices while delivering the essential communication functionality, as will be important for low-power applications such as RF identification (RFID) and wireless sensor networks (WSNs) (see \cite{Grover2010,RZhang2013} and the references therein).

On the other hand, privacy and authentication have increasingly become major concerns for wireless communications and physical (PHY)-layer security has emerged as a new layer of defence to realize perfect secrecy transmission in addition to the costly upper-layer techniques such as cryptography. In this regard, relay-assisted secure transmission was proposed \cite{Tekin2008,Lai2008} and PHY-layer security enhancements by means of cooperative communications have since attracted much attention \cite{Dong2010,JZhang2010,Yang2013,Petropulu2011,Jeong2011,Goel08,Zheng2011,Luo2013,Xing2015,Krikidis2009,Ding2011,Yang2013SPAWC,Li2015,Huang2011}.

In particular, cooperative schemes can be mainly classified into three categories: {\em decode-and-forward (DF)}, {\em amplify-and-forward (AF)}, and {\em cooperative jamming (CJ)} \cite{Dong2010} with CJ being most relevant to PHY-layer security. Specifically, coordinated CJ refers to the scheme of generating a common jamming signal across all single-antenna relay helpers against eavesdropping \cite{Dong2010,Goel08,Zheng2011,Petropulu2011}, while uncoordinated CJ considers that each relay helper emits independent artificial noise (AN) to confound the eavesdroppers \cite{Luo2013,Xing2015}. It is expected that in the scenarios where the direct link is broken between the transmitter (Tx) and the legitimate receiver (Rx), some of the relays have to take on their conventional role of forwarding the information while others will perform CJ \cite{Krikidis2009,Ding2011}. A recent paradigm that generalizes all the above-mentioned cooperation strategies is {\em cooperative beamforming (CB) mixed with CJ} \cite{Yang2013SPAWC,Li2015}, where the available power at each relay is split into two parts: one for forwarding the confidential message and the other for CJ.

However, mixed CB-CJ approaches may be prohibitive in applications with low power devices because idle relays with limited battery supplies would likely prefer saving power for their own traffic to assisting others' communication. In light of this, SWIPT provides the incentive for potential helpers to perform dedicated CB mixed with CJ at no expense of its own power, but opportunistically earn harvested energy. Motivated by this, our work considers secrecy transmission from a Tx to a legitimate Rx with the aid of a set of single-antenna wireless energy harvesting (WEH)-enabled AF-operated relays in the presence of multiple single-antenna eavesdroppers. As a matter of fact, cooperative schemes that involve WEH-enabled relays was recently investigated  in \cite{Ishibashi2014}. We consider the use of the {\em dynamic power splitting (DPS)} receiver architecture, initially proposed for SWIPT in \cite{Zhou2013}, which divides the received power with an adjustable ratio for energy harvesting (EH) and information receiving (IR). WEH-enabled relays using DPS receivers have also been considered in \cite{Nasir2013,Ding2014,Timotheou2014} and \cite{Shi2014,Zhao2015,Ng2014}, without (w/o) and with secrecy consideration, respectively. Note that there is also interest in addressing the threat that WEH receivers may attempt to intercept the confidential messages in SWIPT-enabled networks \cite{Ng2014,Liu2014,Li2014,Xing2014}. Nevertheless, we will focus on exploiting the benefits of WEH-enabled relays when they are trustful.

In particular, motivated by the strong interest in SWIPT and the vast degree-of-freedom (DoF) achievable by cooperative relays, this paper aims to maximize the secrecy rate with the aid of WEH-enabled AF-operated relays, subject to the EH power constraints of individual relays by jointly optimizing the CB of the relays and the CJ covariance matrix.\footnote{The scenario is applicable to WSNs, e.g., a remote health system where a moving patient reports its physical data to a health centre with the aid of intermediary sensor nodes installed on other patients in the vicinity.} In this paper, we assume that there is no direct link between the source and destination nodes, and perfect global channel state information (CSI) is available for the case of centralized optimization.

It is worth pointing out that although our setting may look similar to \cite{Li2015}, their optimal CB-CJ design is not applicable to ours due to the multiplicative nature in beamforming weights incurred by the power splitting (PS) ratios that intrinsically poses more intractability to our optimization problem. Further, our work also differs from \cite{Zhao2015} where an efficient algorithm was proposed to maximize the secrecy rate for the optimization of the PS ratios and AF relay beamforming. The difference is twofold. First, AN was not considered in the second transmission phase in \cite{Zhao2015} and in addition, their algorithm only converged to a local optimum, as opposed to our work that gives the global optimal solutions for CB.

The rest of the paper is organized as follows. Section~\ref{sec:System Model} describes two types of WEH-enabled Rx architecture for the AF relays and defines the secrecy rate region of the relay wiretap channel. Section~\ref{sec:Problem Formulation} then formulates the secrecy rate maximization problems that jointly optimize the AN (or CJ) and the AF-relay CB for the WEH-enabled relays operating with the two types of Rx. The problems are respectively solved by centralized schemes in Section~\ref{sec:centralized} and distributed approaches in Section~\ref{sec:distributed}. Section~\ref{sec:Numerical Results} provides simulation results to evaluate the performance of the proposed schemes. Finally, Section~\ref{sec:conclusion} concludes the paper.

{\em Notations}---We use the uppercase boldface letters for matrices and lowercase boldface letters for vectors. The superscripts $(\cdot)^{T}$, $(\cdot)^{\dag}$, $(\cdot)^{H}$ and $(\cdot)^\ast$ represent, respectively, the transpose, conjugate, conjugate transpose operations on vectors or matrices, and the optimum. In addition, ${\rm trace}(\cdot)$ stands for the trace of a square matrix. Moreover, $[\cdot]_{i,j}$ denotes the $(i,j)$th entry of a matrix, while $\|\cdot\|$ and $\|\cdot\|.^2$ represent the Euclidean norm and the entry-wise absolute value square of a vector, respectively. Also, \({\rm diag}(\cdot)\) denotes a diagonal matrix with its diagonal specified by the given vector and \([\cdot]_{i=1}^N\) represents an $N\times1$ vector with each element indexed by $i$. Furthermore, $\cdot$ and $\circ$ stand for product and Hadamard product, respectively. $\mathbb{C(R)}^{x \times y}$ denotes the field of complex (real) matrices with dimension $x\times y$ and ${\sf E[\cdot]}$ indicates the expectation operation. Finally, $(x)^+$ is short for $\max(0,x)$.

\section{System Model}\label{sec:System Model}
\begin{figure}[htp]
\begin{center}
 \scalebox{0.45}{\includegraphics*{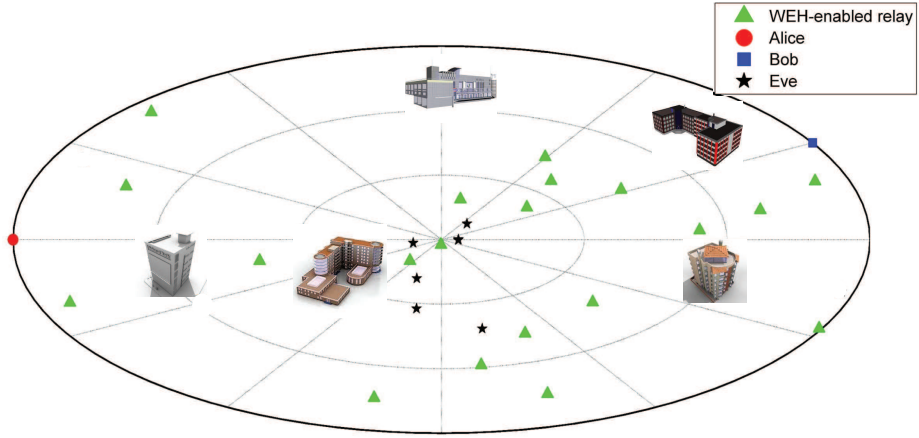}}
 \end{center}
\caption{The system model for an AF relay-assisted SWIPT WSN.}\label{fig:system model}
\end{figure}
In this paper, we consider secrecy transmission in a SWIPT-enabled WSN, where a Tx (Alice) wants to establish confidential communication with the legitimate Rx (Bob) with no direct link but with the aid of $N$ WEH-enabled sensors operating as AF relays, denoted by \(\mathcal{N}=\{1,2,\ldots,N\}\), in the presence of multiple eavesdroppers (Eves), denoted by \(\mathcal{K}=\{1,2,\ldots,K\}\), all equipped with single antenna. We assume that there is no direct link from the Tx to any of the Eves,\footnote{Note that if there exist direct links, our problem formulation and solutions are still applicable without much modification by incorporating destination-aided AN in the first transmit-slot (see \cite{Zhao2015}).} due to, for instance, severe path loss or shadowing, as illustrated in Fig.~\ref{fig:system model}.

We consider a two-hop relaying protocol based on two equal time slots and the duration of one transmit-slot is normalized to be one unit so that the terms ``energy'' and ``power'' are interchangeable with respect to (w.r.t.)~one transmit-slot.

\begin{figure}[htp]
\centering
\subfigure[WEH-enabled relay with SPS.]{ \label{fig:subfig:SPS Rx}\includegraphics[width=3.45in]{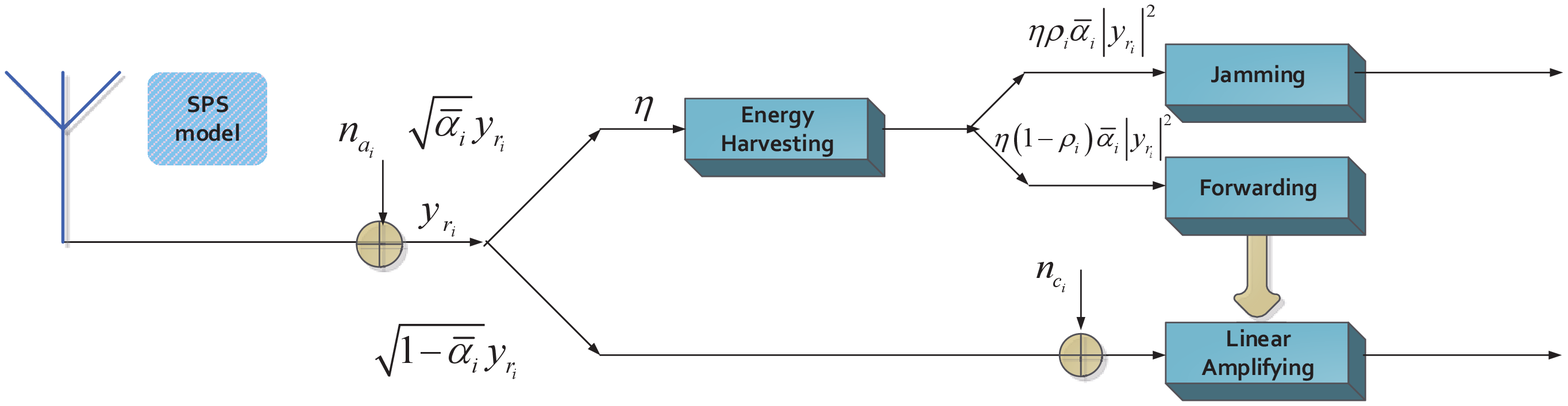}}
\subfigure[WEH-enabled relay with DPS.]
{\label{fig:subfig:DPS Rx}\includegraphics[width=3.45in]{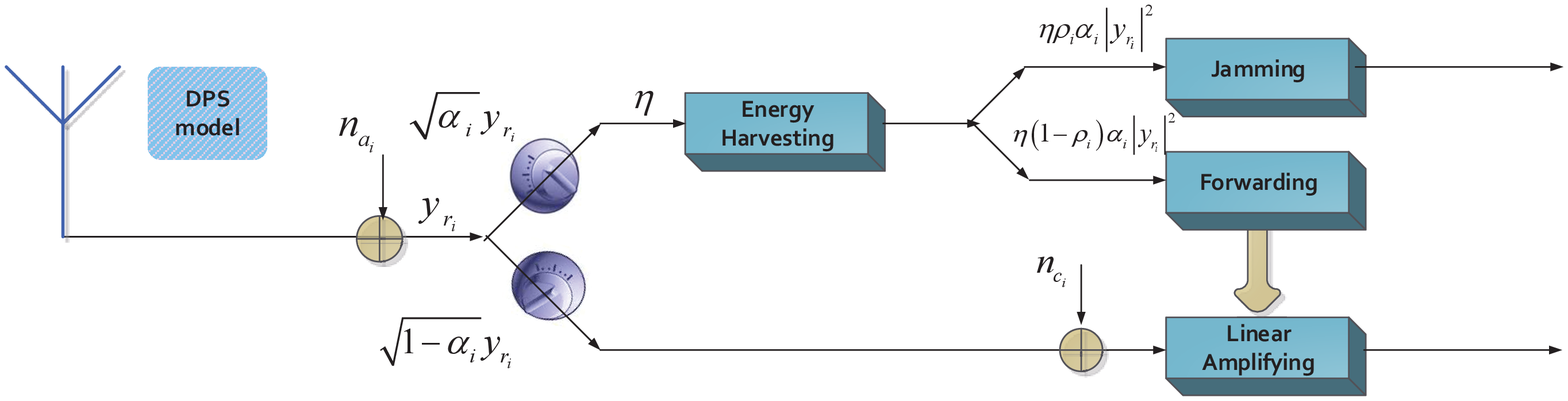}}
\caption{Architectures of the receiver for WEH-enabled relay.}\label{fig:Rx models}
\end{figure}

At the receiver of each AF relay, we introduce two types of WEH-enabled receiver architecture, namely, {\em static power splitting (SPS)} (Fig.~\ref{fig:subfig:SPS Rx}) and {\em DPS} (Fig.~\ref{fig:subfig:DPS Rx}), both of which allow the relay to harvest energy and receive information from the same received signal. Specifically, the receiver first splits a portion of \(\alpha_i\), of the received power for EH and the rest \(1-\alpha_i\) for IR, \(\forall i\). The \(\alpha_i\) portion of harvested power is further divided into two streams with \(\eta\rho_i\alpha_i\vert y_{r_i}\vert^2\) used for generating the AN to confound Eves and \(\eta(1-\rho_i)\alpha_i\vert y_{r_i}\vert^2\) used for amplifying the received signal, where \(y_{r_i}\) is the \(i\)th element of the received signal \(\boldsymbol y_r\in\mathbb{C}^{N\times1}\), and \(0\le\eta<1\) denotes the EH efficiency. Note that DPS with adjustable \(\alpha_i\)'s is presently the most general receiver operation because practical circuits cannot directly decode the information from the stream used for EH \cite{Zhou2013} and SPS is just a special case of DPS with \(\alpha_i=\bar\alpha_i\), \(\forall i\), fixed for the whole transmission duration. However, SPS, advocated for its ease of implementation, is introduced separately in the sequel for its simplified relay beamforming design.

In the first transmit-slot, the received signal at each individual relay can be expressed as
\begin{equation}\label{eq:received at each relay}
y_{r_i}=h_{sr_i}\sqrt{P_s}s+n_{a,i},\; \forall i,
\end{equation}
where the transmit signal \(s\) is a circularly symmetric complex Gaussian (CSCG) random variable with zero mean and unit variance, denoted by \(s\sim\mathcal{CN}(0,1)\), \(h_{sr_i}\) denotes the complex channel from the Tx to the \(i\)th relay, \(P_s\) is the transmit power at the Tx, and \(n_{a,i}\) is the additive white Gaussian noise (AWGN) introduced by the receiving antenna of the \(i\)th relay, denoted by \(n_{a,i}\sim\mathcal{CN}(0,\sigma_{n_a}^2)\).
As such, the linearly amplified baseband equivalent signal at the output of the \(i\)th relay is given by
\begin{equation}\label{eq:forwarded signal at each relay}
x_{r_i1}=\beta_i(\sqrt{1-\alpha_i}y_{r_i}+n_{c,i}), \; \forall i,
\end{equation}
where \(\beta_i\) denotes the complex AF coefficient, and \(n_{c,i}\) denotes the noise due to signal conversion from the RF band to baseband, denoted by \(n_{c,i}\sim\mathcal{CN}(0,\sigma_{n_c}^2)\). Since \(x_{r_i1}\) is constrained by the portion of the harvested power for forwarding, i.e., \(\eta(1-\rho_i)\alpha_i\vert y_{r_i}\vert^2\), \(\beta_i\) is accordingly given by
\begin{equation}\label{eq:amplifying coefficient}
\beta_i=\sqrt{\frac{\eta(1-\rho_i)\alpha_i\vert h_{sr_i}\vert^2P_s}{(1-\alpha_i)\vert h_{sr_i}\vert^2P_s+(1-\alpha_i)\sigma_{n_a}^2+\sigma_{n_c}^2}}e^{j\measuredangle\beta_i},
\end{equation}
where \(\measuredangle\beta_i\) denotes the phase of the AF coefficient for the \(i\)th relay.

Next, we introduce the CJ scheme. Denote the CJ signal generated from $N$ relays by \(\boldsymbol x_{r2}=[x_{r_12},\ldots,x_{r_N2}]^T\) and define its covariance matrix as \(\boldsymbol S={\sf E}[\boldsymbol x_{r2}\boldsymbol x_{r2}^H]\). Then the coordinated CJ transmission can be uniquely determined by the truncated eigenvalue decomposition (EVD) of \(\boldsymbol S\) given by \(\boldsymbol S=\tilde{\boldsymbol V}\tilde{\boldsymbol\Sigma}\tilde{\boldsymbol V}^H\), where \(\boldsymbol\Sigma={\rm diag}([\sigma_1,\ldots,\sigma_d])\) is a diagonal matrix with \(\sigma_j\)'s denoting all the positive eigenvalues of \(\boldsymbol S\) and \(\boldsymbol V\in\mathbb{C}^{N\times d}\) is the precoding matrix satisfying \(\boldsymbol V^H\boldsymbol V=\boldsymbol I\). Note that \(d\le N\) denotes the rank of \(\mv S\) which will be designed later. As a result, the CJ signal can be expressed as
\begin{equation}\label{eq:CJ signal}
\boldsymbol x_{r2}=\sum_{j=1}^d\sqrt{\sigma_j}\boldsymbol v_js_j^\prime,
\end{equation}
where \(\boldsymbol v_j\)'s are drawn from the columns of \(\boldsymbol V\),  and \(s_j^\prime\)'s are independent and identically distributed (i.i.d.) complex Gaussian variables denoted by \(s_j^\prime\sim\mathcal{CN}(0,1)\), which is known as the optimal distribution for AN \cite{Goel08}. On the other hand, \(\vert x_{r_i2}\vert^2\le\eta\rho_i\alpha_i\vert y_{r_i}\vert^2\), \(\forall i\), denotes the power constraint for jamming at the \(i\)th relay, which implies that
\begin{equation}\label{eq:per-relay jamming constraint}
{\rm trace}(\boldsymbol S\boldsymbol E_i)\le\eta\rho_i\alpha_i P_s\vert h_{sr_i}\vert^2, \; \forall i,
\end{equation}
where \(\boldsymbol E_i\) is a diagonal matrix with its diagonal \(\boldsymbol e_i\) (a unit vector with the \(i\)th entry equal to \(1\) and the rest equal to \(0\)).

Note that the CJ scheme proposed above is of the most general form. For the special case when \(d=1\), i.e., \(\boldsymbol x_{r2}=\sqrt{\sigma_1}\boldsymbol v_1s_1^\prime\), each relay transmits a common jamming signal \(s_1^\prime\) with their respective weight drawn from \(\boldsymbol v_1\)  \cite{Dong2010,Zheng2011}. This case is desirable in practice since it has the lowest complexity for implementation.
In summary, the transmitted signal at the \(i\)th relay is given by
\begin{equation}\label{eq:mixed-transmit at each relay}
x_{r_i}=x_{r_i1}+x_{r_i2}, \; \forall i.
\end{equation}

According to \eqref{eq:mixed-transmit at each relay} together with \eqref{eq:received at each relay}, \eqref{eq:forwarded signal at each relay}, and \eqref{eq:CJ signal}, the transmit signal from all relays can be expressed in vector form as
\begin{equation}\label{eq:transmit vector}
\boldsymbol x_r=\boldsymbol D_{\beta\alpha}\boldsymbol h_{sr}\sqrt{P_s}s+\boldsymbol D_{\beta\alpha}\boldsymbol n_a+\boldsymbol D_{\beta}\boldsymbol n_c+\sum_{j=1}^d\sqrt{\sigma_j}\boldsymbol v_js_j^\prime,
\end{equation}
where \(\boldsymbol D_{\beta\alpha}\) and \(\boldsymbol D_{\beta}\) are, respectively, diagonal matrices with their diagonals composed of \((\beta_1\sqrt{1-\alpha_1}, \ldots,\)  \(\beta_N\sqrt{1-\alpha_N})^T\) and \((\beta_1, \ldots, \beta_N)^T\). In addition, \(\boldsymbol h_{sr}=[h_{sr_i}]_{i=1}^N\), \(\boldsymbol n_a=[n_{a,i}]_{i=1}^N\), and \(\boldsymbol n_c=[n_{c,i}]_{i=1}^N\).

\begin{figure*}
\begin{align}
{\rm SINR}_{\rm S,D}&=
\frac{P_s|\boldsymbol h_{rd}^T\boldsymbol D_{\beta\alpha}\boldsymbol h_{sr}|^2}{{\rm trace}(\boldsymbol S\boldsymbol h_{rd}^\dag\boldsymbol h_{rd}^T)+\sigma_{n_a}^2\|\boldsymbol h_{rd}^T\boldsymbol D_{\beta\alpha}\|^2+\sigma_{n_c}^2\|\boldsymbol h_{rd}^T\boldsymbol D_\beta\|^2+\sigma_{n_d}^2}\label{eq:SINR between the Tx and Bob}\\
{\rm SINR}_{\rm S,E,k}&=
\frac{P_s|\boldsymbol h_{re,k}^T\boldsymbol D_{\beta\alpha}\boldsymbol h_{sr}|^2}{{\rm trace}(\boldsymbol S\boldsymbol h_{re,k}^\dag\boldsymbol h_{re,k}^T)+\sigma_{n_a}^2\|\boldsymbol h_{re,k}^T\boldsymbol D_{\beta\alpha}\|^2+\sigma_{n_c}^2\|\boldsymbol h_{re,k}^T\boldsymbol D_\beta\|^2+\sigma_{n_e,k}^2}\label{eq:SINR between the Tx and the kth Eve}
\end{align}
\hrulefill
\end{figure*}

In the second transmit-slot,  the received signal at the desired receiver, i.e., Bob, is given by
\begin{equation}\label{eq:received at the Rx}
y_d =\boldsymbol h_{rd}^T\boldsymbol x_r+n_d,
\end{equation}
where \(\boldsymbol h_{rd}=[h_{r_id}]_{i=1}^N\) comprises complex channels from the \(i\)th relay to the Rx and \(n_d\sim\mathcal{CN}(0,\sigma_{n_d}^2)\) is the corresponding receiving AWGN. By substituting \eqref{eq:transmit vector} into \eqref{eq:received at the Rx}, \(y_d\) can be expressed as
\begin{align}
 y_d=\boldsymbol h_{rd}^T\boldsymbol D_{\beta\alpha}\boldsymbol h_{sr}\sqrt{P_s}s+\boldsymbol h_{rd}^T\boldsymbol D_{\beta\alpha}\boldsymbol n_a+\boldsymbol h_{rd}^T\boldsymbol D_{\beta}\boldsymbol n_c
+\boldsymbol h_{rd}^T\sum_{j=1}^d\sqrt{\sigma_j}\boldsymbol v_js_j^\prime+n_d.
\end{align}
The received signal at the \(k\)th Eve, \(k\in\mathcal{K}\), is given by
\begin{align}
y_{e,k}=\boldsymbol h_{re,k}^T\boldsymbol D_{\beta\alpha}\boldsymbol h_{sr}\sqrt{P_s}s+\boldsymbol h_{re,k}^T\boldsymbol D_{\beta\alpha}\boldsymbol n_a
+\boldsymbol h_{re,k}^T\boldsymbol D_{\beta}\boldsymbol n_c+\boldsymbol h_{re,k}^T\sum_{j=1}^d\sqrt{\sigma_j}\boldsymbol v_js_j^\prime+n_{e,k},
\end{align}
where \(\boldsymbol h_{re,k}=[h_{r_ie,k}]_{i=1}^N\) denotes the complex channels from the relays to the \(k\)th Eve and \(n_e\sim\mathcal{CN}(0,\sigma_{n_e}^2)\) is the AWGN at the \(k\)th eavesdropper.

The mutual information for the  Rx (Bob) is given by \(r_{\rm S,D}=\tfrac{1}{2}\log_2(1+{\rm SINR}_{\rm S,D})\), and that for the \(k\)th Eve is \(r_{\rm S,E,k}=\tfrac{1}{2}\log_2(1+{\rm SINR}_{\rm S,E,k})\), \(\forall k\), where \({\rm SINR}_{\rm S,D}\) and \({\rm SINR}_{\rm S,E,k}\), which denote their respective signal-to-interference-plus-noise ratios (SINRs), are given at the top of next page.

Next, we define the secrecy rate region that consists of all the achievable secrecy rate for the relay wiretap channel given transmit power \(P_s\), denoted by \(\mathcal{R}(\{\measuredangle\beta_i\}, \{\rho_i\}, \{\alpha_i\},\boldsymbol S)\), which is given by \cite{Liang2009,Dong2010}
\begin{align}\label{eq:secrecy rate region}
\mathcal{R}(\{\measuredangle\beta_i\} \{\rho_i\}, \{\alpha_i\},\boldsymbol S)\triangleq\!\!\bigcup_{\{\measuredangle\beta_i\},\{\rho_i\},\{\alpha_i\},\eqref{eq:per-relay jamming constraint}}\!\!\left\{r_{\rm sec}:
r_{\rm sec}\le\left(r_{\rm S,D}-\max\limits_{k\in\mathcal{K}}~r_{\rm S,E,k}\right)^+\right\}.
\end{align}

\section{Problem Formulation}\label{sec:Problem Formulation}
\subsection{AN-Aided Secrecy Relay Beamforming for SPS}\label{sec:problem formulation for SPS}
In this section, we consider the secrecy rate maximization problem by jointly optimizing the AN beams, relay beam and their power allocations for WEH-enabled AF relays operating with SPS, i.e., \(\alpha_i=\bar\alpha_i\), \(\forall i\), is fixed.

By replacing \(\beta_i\) with \eqref{eq:amplifying coefficient}, \(|\boldsymbol h_{rd}^T\boldsymbol D_{\beta\alpha}\boldsymbol h_{sr}|^2\) in \eqref{eq:SINR between the Tx and Bob} at the next page becomes
\begin{equation}\label{eq:numerator}
|\boldsymbol h_{rd}^T\boldsymbol D_{\beta\alpha}\boldsymbol h_{sr}|^2=\left|\sum_{i=1}^Nw_{1,i}[\tilde{\boldsymbol h}_{sd}]_i\right|^2,
\end{equation}
where \(w_{1,i}=\sqrt{1-\rho_i}e^{j\measuredangle\beta_i}\) and
\begin{equation}
[\tilde{\boldsymbol h}_{sd}]_i\triangleq h_{sr_i}h_{r_id}\sqrt{\tfrac{\eta\bar\alpha_i(1-\bar\alpha_i)|h_{sr_i}|^2P_s}{(1-\bar\alpha_i)(|h_{sr_i}|^2P_s
+\sigma_{n_a}^2)+\sigma_{n_c}^2}}.
\end{equation}
In addition, \(\sigma_{n_a}^2\|\boldsymbol h_{rd}^T\boldsymbol D_{\beta\alpha}\|^2\) and \(\sigma_{n_c}^2\|\boldsymbol h_{rd}^T\boldsymbol D_\beta\|^2\) in \eqref{eq:SINR between the Tx and Bob} can also be combined as
\begin{equation}\label{eq:denominator}
\sigma_{n_a}^2\|\boldsymbol h_{rd}^T\boldsymbol D_{\beta\alpha}\|^2+\sigma_{n_c}^2\|\boldsymbol h_{rd}^T\boldsymbol D_\beta\|^2=\sum_{i=1}^N|w_{1,i}|^2[\boldsymbol D_{\hat{sd}}]_{i,i},
\end{equation}
where
\begin{align}
[\boldsymbol D_{\hat{sd}}]_{i,i}=\frac{\eta\bar\alpha_iP_s|h_{sr_i}|^2|h_{r_id}|^2((1-\bar\alpha_i)\sigma_{n_a}^2+\sigma_{n_c}^2)}
{(1-\bar\alpha_i)(|h_{sr_i}|^2P_s+\sigma_{n_a}^2)+\sigma_{n_c}^2}.
\end{align}
As a result, \(r_{\rm S,D}\) can be rewritten as
\begin{equation}\label{eq:compact IMI for the MU with CJ for SPS}
  r_{\rm S,D}=\frac{1}{2}\log_2\left(1+\tfrac{P_s|\tilde{\boldsymbol h}_{sd}^T\boldsymbol w_1|^2}{{\rm trace}(\boldsymbol S\boldsymbol h_{rd}^\dag\boldsymbol h_{rd}^T)+\boldsymbol w_1^H\boldsymbol D_{\hat{sd}}\boldsymbol w_1+\sigma_{n_d}^2}\right),
\end{equation}
where \(\boldsymbol w_1=[w_{1,i}]_{i=1}^N\). Similarly by letting
\begin{equation}
[\tilde{\boldsymbol h}_{se,k}]_i\triangleq h_{sr_i}h_{r_ie,k}\sqrt{\frac{\eta\bar\alpha_i(1-\bar\alpha_i)|h_{sr_i}|^2P_s}{(1-\bar\alpha_i)(|h_{sr_i}|^2P_s
+\sigma_{n_a}^2)+\sigma_{n_c}^2}}
\end{equation}
and
\begin{equation}\label{eq:D_hat_se}
[\boldsymbol D_{\hat{se},k}]_{i,i}\triangleq\frac{\eta\bar\alpha_iP_s|h_{sr_i}|^2|h_{r_ie,k}|^2((1-\bar\alpha_i)\sigma_{n_a}^2+\sigma_{n_c}^2)}
{(1-\bar\alpha_i)(|h_{sr_i}|^2P_s+\sigma_{n_a}^2)+\sigma_{n_c}^2},
\end{equation}
we have
\begin{equation}\label{eq:compact IMI for the kth Eve with CJ for SPS}
r_{\rm S,E,k}=\frac{1}{2}\log_2\left(1+\tfrac{P_s|\tilde{\boldsymbol h}_{se,k}^T\boldsymbol w_1|^2}{{\rm trace}(\boldsymbol S\boldsymbol h_{re,k}^\dag\boldsymbol h_{re,k}^T)+\boldsymbol w_1^H\boldsymbol D_{\hat{se},k}\boldsymbol w_1+\sigma_{n_e,k}^2}\right).
\end{equation}

By some simple manipulation, \eqref{eq:per-relay jamming constraint} is reformulated as a per-relay jamming power  constraint given by
\begin{equation}\label{eq:per-relay jamming constraint for SPS}
{\rm trace}(\boldsymbol S\boldsymbol E_i)\le\eta\bar\alpha_i P_s\vert h_{sr_i}\vert^2(1-\vert w_{1,i}\vert^2), \; \forall i.
\end{equation}

Now, the secrecy rate maximization problem w.r.t.~\(\rho_i\)'s, \(\measuredangle\beta_i\)'s and \(\boldsymbol S\) for SPS-based relays can be formulated as
\begin{equation*}
\mathrm{(P1)}:\max_{\boldsymbol{w}_1,\boldsymbol{S}}~
\left(\eqref{eq:compact IMI for the MU with CJ for SPS}-\max\limits_{k\in\mathcal{K}}\eqref{eq:compact IMI for the kth Eve with CJ for SPS}\right)^+~{\rm s.t.}~\eqref{eq:per-relay jamming constraint for SPS}, \ \mv S\succeq \mv 0.
\end{equation*}

\begin{figure*}
\begin{align}
{\rm SINR}_{\rm S, D}&=\frac{P_s|\boldsymbol s_{sd}^T\boldsymbol u_1|^2}{{\rm trace}(\boldsymbol S\boldsymbol h_{rd}^\dag\boldsymbol h_{rd}^T)+\sigma_{n_a}^2\boldsymbol u_1^H {\rm diag}(\boldsymbol c_{0}\circ\|\boldsymbol h_{rd}\|.^2)\boldsymbol u_1+\sigma_{n_c}^2\boldsymbol u_2^H {\rm diag}(\boldsymbol c_{0}\circ\|\boldsymbol h_{rd}\|.^2) \boldsymbol u_2+\sigma_{n_d}^2}\label{eq:SINR for the MU with CJ for DPS}\\
{\rm SINR}_{\rm S, E,k}&=\frac{P_s|\boldsymbol s_{se,k}^T\boldsymbol u_1|^2}{{\rm trace}(\boldsymbol S\boldsymbol h_{re,k}^\dag\boldsymbol h_{re,k}^T)+\sigma_{n_a}^2\boldsymbol u_1^H {\rm diag}(\boldsymbol c_{0}\circ\|\boldsymbol h_{re,k}\|.^2)\boldsymbol u_1+\sigma_{n_c}^2\boldsymbol u_2^H {\rm diag}(\boldsymbol c_{0}\circ\|\boldsymbol h_{re,k}\|.^2) \boldsymbol u_2+\sigma_{n_e,k}^2}\label{eq:SINR for the kth Eve with CJ for DPS}
\end{align}
\hrulefill
\end{figure*}

\subsection{AN-Aided Secrecy Relay Beamforming for DPS}\label{sec:problem formulation for DPS}
Here, we consider the secrecy rate maximization problem for WEH-enabled AF relays with adjustable PS ratios \(\{\alpha_i\}\) by jointly optimizing the AN beams, relay beam, WEH PS ratios \(\{\alpha_i\}\), and AN PS ratios \(\{\rho_i\}\).

First, consider the following variable transformation:
\begin{equation}\label{eq:u1 and u2}
 \left\{\begin{array}{l}
 u_{1,i}=\sqrt{\frac{\alpha_i(1-\alpha_i)(1-\rho_i)}{(1-\alpha_i)(|h_{sr_i}|^2P_s
+\sigma_{n_a}^2)+\sigma_{n_c}^2}}e^{j\measuredangle\beta_i} \\
u_{2,i}=\sqrt{\frac{\alpha_i(1-\rho_i)}{(1-\alpha_i)(|h_{sr_i}|^2P_s
+\sigma_{n_a}^2)+\sigma_{n_c}^2}} \end{array}, \; \forall i.\right.
\end{equation}
Using this, \(\vert \boldsymbol h_{rd}^T\boldsymbol D_{\beta\alpha}\boldsymbol h_{sr}\vert^2\) can then be expressed as \(\vert \boldsymbol s_{sd}^T\boldsymbol u_1\vert^2\), where \(\boldsymbol s_{sd}=[h_{sr_i}h_{r_id}\sqrt{\eta\vert h_{sr_i}\vert^2 P_s}]_{i=1}^N\) and \(\boldsymbol u_1=[u_{1,i}]_{i=1}^N\). Moreover, \(\|\boldsymbol h_{rd}^T\boldsymbol D_{\beta\alpha}\|^2\) and \(\|\boldsymbol h_{rd}^T\boldsymbol D_\beta\|^2\) can be simplified as \(\boldsymbol u_1^H {\rm diag}(\boldsymbol c_{0}\circ\|\boldsymbol h_{rd}\|.^2)\boldsymbol u_1\) and \(\boldsymbol u_2^H {\rm diag}(\boldsymbol c_{0}\circ\|\boldsymbol h_{rd}\|.^2)\boldsymbol u_2\), respectively, where \(\boldsymbol c_{0}=[c_{0,i}]_{i=1}^N\) with \(c_{0,i}=\eta P_s\vert h_{sr_i}\vert^2\), \(\forall i\), and \(\boldsymbol u_2=[u_{2,i}]_{i=1}^N\). Similarly, we have \(\vert \boldsymbol h_{re,k}^T\boldsymbol D_{\beta\alpha}\boldsymbol h_{sr}\vert^2=\vert \boldsymbol s_{se,k}^T\boldsymbol u_1\vert^2\), \(\|\boldsymbol h_{re,k}^T\boldsymbol D_{\beta\alpha}\|^2=\boldsymbol u_1^H {\rm diag}(\boldsymbol c_{0}\circ\|\boldsymbol h_{re,k}\|.^2)\boldsymbol u_1\), and \(\|\boldsymbol h_{re,k}^T\boldsymbol D_\beta\|^2=\boldsymbol u_2^H {\rm diag}(\boldsymbol c_{0}\circ\|\boldsymbol h_{re,k}\|.^2) \boldsymbol u_2\), where \(\boldsymbol s_{se,k}=[h_{sr_i}h_{r_ie,k}\sqrt{\eta\vert h_{sr_i}\vert^2 P_s}]_{i=1}^N\), \(\forall k\in\mathcal{K}\).

Then we apply the above transformation to \({\rm SINR}_{\rm S, D}\) (c.f.~\eqref{eq:SINR between the Tx and Bob}) and \({\rm SINR}_{\rm S, E,k}\) (c.f.~\eqref{eq:SINR between the Tx and the kth Eve}), \(\forall k\), to get \eqref{eq:SINR for the MU with CJ for DPS} and \eqref{eq:SINR for the kth Eve with CJ for DPS} (see next page). Now, we can recast the constraints w.r.t.~\(\boldsymbol S\), \(\alpha_i\)'s, and \(\rho_i\)'s to those w.r.t.~the transformed variables \(u_{1,i}\)'s and \(u_{2,i}\)'s. In accordance with \eqref{eq:u1 and u2}, the optimization variables, \(\alpha_i\)'s and \(\rho_i\)'s, can be alternatively given by
\begin{align}\label{eq:alpha and rho}
 \left\{\begin{array}{l}
 \alpha_i=1-\frac{\vert u_{1,i}\vert^2}{\vert u_{2,i}\vert^2}\\
\rho_i=1-\frac{\vert u_{2,i}\vert^2(c_{1,i}\vert u_{1,i}\vert^2+\sigma_{n_c}^2\vert u_{2,i}\vert^2)}{\vert u_{2,i}\vert^2-\vert u_{1,i}\vert^2} \end{array}\right., \; \forall i,
\end{align}
where \(c_{1,i}=P_s\vert h_{sr_i}\vert^2+\sigma_{n_a}^2\). Replacing \(\alpha_i\)'s and \(\rho_i\)'s with \eqref{eq:alpha and rho}, \eqref{eq:per-relay jamming constraint} is reformulated as
\begin{align}\label{eq:per-relay jamming constraint for DPS}
 {\rm trace}(\boldsymbol S\boldsymbol E_i)\le c_{0,i}\left(1-\frac{\vert u_{2,i}\vert^2(c_{1,i}\vert u_{1,i}\vert^2+\sigma_{n_c}^2\vert u_{2,i}\vert^2)}{\vert u_{2,i}\vert^2-\vert u_{1,i}\vert^2}\right)
 \left(1-\frac{\vert u_{1,i}\vert^2}{\vert u_{2,i}\vert^2}\right), \; \forall i.
  \end{align}
 On the other hand, since \(\alpha_i\ge 0\) and \(\rho_i\ge 0\), \(\forall i\), after some simple manipulation, it follows from \eqref{eq:alpha and rho} that
 \begin{align}
 \!\!   & \vert u_{1,i}\vert^2-\vert u_{2,i}\vert^2\le 0, \; \forall i, \label{eq:constraint on alpha} \\
 \!\!   &  \vert u_{2,i}\vert^2(c_{1,i}\vert u_{1,i}\vert^2+\sigma_{n_c}^2\vert u_{2,i}\vert^2)\le \vert u_{2,i}\vert^2-\vert u_{1,i}\vert^2, \; \forall i.  \label{eq:constraint on rho}
 \end{align}

As such, the secrecy rate maximization problem for DPS-based relays becomes
\begin{align*}
\mathrm{(P2)}: \max_{\boldsymbol{u}_1,\boldsymbol{u}_2,\boldsymbol{S}}
& ~~~\Big(\frac{1}{2}\log_2(1+{\rm SINR}_{\rm S,D})\notag\\
&~~~~-\frac{1}{2}\log_2(1+\max\limits_{k\in\mathcal{K}}{\rm SINR}_{\rm S,E,k})\Big)^+ \\
{\rm s.t.}& ~~~\eqref{eq:per-relay jamming constraint for DPS}, \, \eqref{eq:constraint on alpha}, \, \mbox{and} \, \eqref{eq:constraint on rho}.
\end{align*}

\section{Centralized Secure AF Relaying}\label{sec:centralized}
In this section, we resort to centralized approaches to solve problem \(\mathrm{(P1)}\) and \(\mathrm{(P2)}\), respectively, assuming that there is a central optimizer that is able to collect all CSIs including \(\boldsymbol h_{sr}\), \(\boldsymbol h_{rd}\) and \(\boldsymbol h_{re}\), perform the optimization, and broadcast to relays their individual optimized parameters.

\subsection{Optimal Solutions for SPS}\label{sec:optimal solutions for SPS}
To start with, we recast \(\mathrm{(P1)}\) into a two-stage problem by introducing a slack variable \(\tau\). First of all, we solve the epigraph reformulation of  \(\mathrm{(P1)}\) with a fixed \(\tau\in(0,1]\) as
\begin{align*}
\mathrm{(P1.1)}:&\max_{\boldsymbol{w}_1,\boldsymbol{S}\succeq {\bf 0}} ~~
\frac{P_s|\tilde{\boldsymbol h}_{sd}^T\boldsymbol w_1|^2}{{\rm trace}(\boldsymbol S\boldsymbol h_{rd}^\dag\boldsymbol h_{rd}^T)+\boldsymbol w_1^H\boldsymbol D_{\hat{sd}}\boldsymbol w_1+\sigma_{n_d}^2}\notag\\
&{\rm s.t.} ~~\eqref{eq:per-relay jamming constraint for SPS}~{\rm and}\notag\\
&1+\frac{P_s|\tilde{\boldsymbol h}_{se,k}^T\boldsymbol w_1|^2}{{\rm trace}(\boldsymbol S\boldsymbol h_{re,k}^\dag\boldsymbol h_{re,k}^T)+\boldsymbol w_1^H\boldsymbol D_{\hat{se,k}}\boldsymbol w_1+\sigma_{n_e,k}^2}\notag\\
&~~~~~~~~~~~~~~~\le1/\tau,\; \forall k.
\end{align*}

Defining \(f_1(\tau)\) as the optimum value of \(\mathrm{(P1.1)}\) and denoting \(H_1(\tau)=\tau f_1(\tau)\), the objective function of \(\mathrm{(P1)}\) is given by
\begin{align}\label{eq:obj function of (P1.2)}
\frac{1}{2}\log_2(1+f_1(\tau))-\frac{1}{2}\log_2(1/\tau)=\frac{1}{2}\log_2(\tau+H_1(\tau)),
\end{align}
where \((\cdot)^+\) in the objective function has been omitted and we claim a zero secrecy rate if \eqref{eq:obj function of (P1.2)} admits a negative value. As a result, \(\mathrm{(P1)}\) can be equivalently given by
\begin{equation*}
\mathrm{(P1.2)}:\max_{\tau_{\min,1}\le\tau\le 1}\log_2(\tau+H_1(\tau)).
\end{equation*}
Note that this single-variable optimization problem allows for simple one-dimension search over \(\tau\in[\tau_{\min,1},1]\), assuming that \(H_1(\tau)\) is attainable given any \(\tau\) in this region. As the physical meaning of \(1/\tau-1\) in \(\mathrm{(P1.1)}\) can be interpreted as the maximum permitted SINR for the best eavesdropper's channel, feasibility for a non-zero secrecy rate implies that
\begin{align}\label{eq:tau_{min,1}}
\tau&\ge\frac{1}{1+\frac{P_s\vert \tilde{\boldsymbol h}_{sd}^T\boldsymbol w_1\vert^2/\sigma_{n_d}^2}{{\rm trace}(\boldsymbol{S}\boldsymbol h_{rd}^\dag\boldsymbol h_{rd}^T)/\sigma_{n_d}^2+\boldsymbol w_1^H\boldsymbol D_{\hat{sd}}\boldsymbol w_1/\sigma_{n_d}^2+1}}\notag\\
&\stackrel{(a)}{\ge}\frac{1}{1+P_s\|\tilde{\boldsymbol h}_{sd}\|^2\|\boldsymbol w_1\|^2/\sigma_{n_d}^2}\notag\\
&\stackrel{(b)}{\ge}\frac{1}{1+NP_s\|\tilde{\boldsymbol h}_{sd}\|^2/\sigma_{n_d}^2}=\tau_{\min,1},
\end{align}
where Cauchy-Schwarz inequality has been applied in \((a)\) and \((b)\) follows from \(\vert w_{1,i}\vert^2\le1\), \(\forall i\in\mathcal{N}\).

The above epigraph reformulation of non-convex problems like  \(\mathrm{(P1)}\) has been widely employed in the literature \cite{Li2013,Li2015}, and \(\mathrm{(P1.2)}\) admits the same optimal value as \(\mathrm{(P1)}\) while \(\mathrm{(P1.1)}\) with the optimal \(\tau\) provides the corresponding optimal solution to \(\mathrm{(P1)}\). We summarize the steps for solving \(\mathrm{(P1)}\) here: given any \(\tau\in[\tau_{\min,1},1]\), solve \(\mathrm{(P1.1)}\) to obtain \(H_1(\tau)\); solve \(\mathrm{(P1.2)}\) via a one-dimensional search over \(\tau\). Before developing solutions to \(\mathrm{(P1.1)}\), we have the lemma below.

\begin{lemma}
\(H_1(\tau)\) is a concave function of \(\tau\). \label{lemma:concavity of H_1(tau)}
\end{lemma}

\begin{IEEEproof}
See Appendix~\ref{appendix:proof of concavity of H_1(tau)}.
\end{IEEEproof}

\begin{remark}
Using Lemma \ref{lemma:concavity of H_1(tau)}, it is easy to verify that \(\frac{1}{2}\log_2(\tau+H_1(\tau))\) is also a concave function of \(\tau\) according to the composition rule \cite[pp.~84]{Boyd2004}, which allows for a more effective search for the optimum \(\tau\), e.g., bi-section method, than the exhaustive search used in \cite{Liu2014}. Moreover, although \(H_1(\tau)\) is not differentiable w.r.t.~\(\tau\), the bi-section method can still be implemented, the algorithm involving which is similarly applied in solving \(\mathrm{(P2)}\) and thus will be given later in Section~\ref{sec:proposed solutions for DPS}.
\end{remark}

In the sequel, we focus on solving \(\mathrm{(P1.1)}\). By introducing \(\boldsymbol X_1=\boldsymbol w_1\boldsymbol w_1^H\) and ignoring the rank-one constraint on \(\boldsymbol X_1\), \( \mathrm{(P1.1)}\) can be alternatively solved by
\begin{align*}
&~\mathrm{(P1.1\text{-}SDR)}:\\
&\left\{\begin{aligned}
&\max_{\boldsymbol X_1,\boldsymbol S\succeq \boldsymbol 0}
~\frac{\tau P_s{\rm trace}(\boldsymbol X_1\tilde{\boldsymbol h}_{sd}^\dag\tilde{\boldsymbol h}_{sd}^T)}{{\rm trace}(\boldsymbol S\boldsymbol h_{rd}^\dag\boldsymbol h_{rd}^T)+{\rm trace}(\boldsymbol X_1\boldsymbol D_{\hat{sd}})+\sigma_{n_d}^2}\\
&{\rm s.t.} ~\frac{P_s{\rm trace}(\boldsymbol X_1\tilde{\boldsymbol h}_{se,k}^\dag\tilde{\boldsymbol h}_{se,k}^T)}{{\rm trace}(\boldsymbol S\boldsymbol h_{re,k}^\dag\boldsymbol h_{re,k}^T)+{\rm trace}(\boldsymbol X_1\boldsymbol D_{\hat{se},k})+\sigma_{n_e,k}^2}\\
&~\hspace{1.89in}\le\frac{1}{\tau}-1,\; \forall k,\\
&{\rm trace}((\boldsymbol S+\eta\bar\alpha_i P_s\vert h_{sr_i}\vert^2\boldsymbol X_1)\boldsymbol E_i)\le\eta\bar\alpha_i P_s\vert h_{sr_i}\vert^2,\; \forall i.
\end{aligned}\right.
\end{align*}
Note that the objective function has been multiplied by \(\tau\) compared with that of \(\mathrm{(P1.1)}\) for ready computation of \(H_1(\tau)\).

Although \(\mathrm{(P1.1\text{-}SDR)}\) is made easier to solve than \(\mathrm{(P1.1)}\) by rank relaxation, it is still a quasi-convex problem considering the linear fractional form of the objective function and constraints, for which Charnes-Cooper transformation \cite{Charnes1962} will be applied for equivalent convex reformulation. Specifically, by substituting \(\boldsymbol X_1=\hat{\boldsymbol X}_1/\xi\) and \(\boldsymbol S=\hat{\boldsymbol S}/\xi\) into \(\mathrm{(P1.1\text{-}SDR)}\), it follows that
\begin{align*}
&~\mathrm{(P1.1\text{-}SDP)}:\\
& \left\{\begin{aligned}
&\max_{\hat{\boldsymbol X}_1,\hat{\boldsymbol S}\succeq \boldsymbol 0,\xi\ge 0}
~P_s{\rm trace}(\hat{\boldsymbol X}_1\tilde{\boldsymbol h}_{sd}^\dag\tilde{\boldsymbol h}_{sd}^T)\\
&{\rm s.t.} ~{\rm trace}(\hat{\boldsymbol S}\boldsymbol h_{rd}^\dag\boldsymbol h_{rd}^T)+{\rm trace}(\hat{\boldsymbol X}_1\boldsymbol D_{\hat{sd}})+\xi\sigma_{n_d}^2=\tau,\\
&\left(\frac{1}{\tau}-1\right)\left({\rm trace}(\hat{\boldsymbol S}\boldsymbol h_{re,k}^\dag\boldsymbol h_{re,k}^T)+{\rm trace}(\hat{\boldsymbol X}_1\boldsymbol D_{\hat{se},k})\right.\\
&~~~~~~~~\left.+\xi\sigma_{n_e,k}^2 \right )
\ge P_s{\rm trace}(\hat{\boldsymbol X}_1\tilde{\boldsymbol h}_{se,k}^\dag\tilde{\boldsymbol h}_{se,k}^T),\; \forall k,\\
&{\rm trace}((\hat{\boldsymbol S}+\eta\bar\alpha_i P_s\vert h_{sr_i}\vert^2\hat{\boldsymbol X}_1)\boldsymbol E_i)\le\xi\eta\bar\alpha_i P_s\vert h_{sr_i}\vert^2,\; \forall i.
\end{aligned}\right.
\end{align*}
Problem \(\mathrm{(P1.1\text{-}SDP)}\) can now be optimally and efficiently solved using interior-point based methods by some off-the-shelf convex optimization toolboxes, e.g., CVX \cite{CVX}.


\begin{proposition}
We have the following results:
\begin{enumerate}[1)]
  \item The optimal solution to \(\mathrm{(P1.1\text{-}SDP)}\) satisfies \({\rm rank}(\hat{\boldsymbol X}_1^\ast)=1\);
 \item \(\hat{\mv X}_1^\ast=\hat{\boldsymbol w}_1^\ast\hat{\boldsymbol w}_1^{\ast H}\), where \(\hat{\boldsymbol w}_1^\ast\) is given by
 \begin{align}
\hat{\boldsymbol w}_1^\ast=\sqrt{\frac{\tau-\xi^\ast\sigma_{n_d}^2-{\rm trace}(\hat{\boldsymbol S}^\ast\boldsymbol h_{rd}^\dag\boldsymbol h_{rd}^T)}{{\rm trace}(\hat{\boldsymbol w}_1\hat{\boldsymbol w}_1^H\boldsymbol D_{\hat{sd}})}}\hat{\boldsymbol w}_1, \label{eq:optimal w1}
 \end{align}
in which \(\hat{\boldsymbol w}_1\) is given in Appendix~\ref{appendix:proof of rank-one solution with CJ for SPS};
\item \({\rm rank}(\hat{\boldsymbol S}^\ast)\le\min(K,N)\).
\end{enumerate}
\label{prop:rank-one solution with CJ for SPS}
\end{proposition}

\begin{IEEEproof}
See Appendix~\ref{appendix:proof of rank-one solution with CJ for SPS}.
\end{IEEEproof}

Proposition~\ref{prop:rank-one solution with CJ for SPS} implies that the rank-one relaxation of \(\mathrm{(P1.1\text{-}SDR)}\) from \(\mathrm{(P1.1)}\) is tight for an arbitrary given \(\tau\). The \(\rho^\ast\)'s and \(\measuredangle\beta_i^\ast\)'s can thus be retrieved from the magnitude and angle of \(\boldsymbol w_1^\ast\), respectively, by applying EVD to \(\boldsymbol X_1^\ast\).

\subsection{Proposed Solutions for DPS}\label{sec:proposed solutions for DPS}
Similar to Section~\ref{sec:optimal solutions for SPS}, in this section, we aim at solving the two-stage reformulation of \(\mathrm{(P2)}\) by introducing a slack variable \(\tau\in[\tau_{\min,2},1]\). First, for a given \(\tau\), we solve
\begin{align*}
\mathrm{(P2.1)}:\max_{\boldsymbol u_1,\boldsymbol u_2,\boldsymbol S}
\eqref{eq:SINR for the MU with CJ for DPS}~~
{\rm s.t.}~~\eqref{eq:SINR for the kth Eve with CJ for DPS}, \, \forall k, \; \eqref{eq:per-relay jamming constraint for DPS}-\eqref{eq:constraint on rho}.
\end{align*}
Next, denoting \(\tau f_2(\tau)\) by \(H_2(\tau)\), where \(f_2(\tau)\) is the optimum value for problem \(\mathrm{(P2.1)}\), we solve the following problem that attains the same optimum value as \(\mathrm{(P2)}\):
\begin{align*}
\mathrm{(P2.2)}:\max_{\tau}\log_2(\tau+H_2(\tau))~~{\rm s.t.}~~\tau_{\min,2}\le\tau\le 1,
\end{align*}
where \(\tau_{\min,2}\) is similarly derived as \(\tau_{\min,1}\) so that we directly arrive at \(\tau\ge\tfrac{1}{1+P_s\|\boldsymbol s_{sd}\|^2\sum_{i=1}^N\frac{1}{\sigma_{n_d}^2(\vert h_{sr_i}\vert^2 P_s+\sigma_{n_a}^2+\sigma_{n_c}^2)}}\), denoted by \(\tau_{\min,2}\). We claim that \(\mathrm{(P2.2)}\) can be solved by bi-section for \(\tau\) over the interval \([\tau_{\min,2},1]\) assuming that \(H_2(\tau)\) is valid for any given \(\tau\) (Otherwise a zero secrecy rate, i.e., \(H_2(\tau)=0\), is returned.), since \(H_2(\tau)\) has the following property.

\begin{lemma}
\(H_2(\tau)\) is a concave function of \(\tau\). \label{lemma:concavity of H_2(tau)}
\end{lemma}

\begin{IEEEproof}
The proof is similar to that for Lemma~\ref{lemma:concavity of H_1(tau)}, and thus is omitted.
\end{IEEEproof}

It is also seen that how to attain \(H_2(\tau)\) forms the main thrust for solving \(\mathrm{(P2)}\). However, the constraints in \eqref{eq:per-relay jamming constraint for DPS}, \eqref{eq:constraint on alpha} and \eqref{eq:constraint on rho} are not convex w.r.t.~\(u_{1,i}\) and/or \(u_{2,i}\), \(\forall i\), due to their high orders and multiplicative structure. \(\mathrm{(P2.1)}\) thus turns out to be very hard to solve in general. To cope with these non-convex constraints, we introduce the following lemma.

\begin{lemma}[\cite{Timotheou2014}]
  The restricted hyperbolic constraints which have the form \(\boldsymbol x^H\boldsymbol x\le yz\), where \(\boldsymbol x\in\mathbb{C}^{N\times1}\), \(y,\ z\ge0\), are equivalent to rotated second-order cone (SOC) constraints as follows.
  \begin{equation}
  \left\|\left(\begin{array}{c}2\boldsymbol x\\ y-z\end{array}\right)\right\|\le y+z.\label{eq:restricted hyperbolic constraints}
  \end{equation}\label{lemma:hyperbolic to SOC}
\end{lemma}

For convenience, denoting \(\vert u_{1,i}\vert^2\), \(\vert u_{2,i}\vert^2\), \({\rm trace}(\boldsymbol S\boldsymbol E_i)\) by \(x_i\), \(y_i\), and \(z_i\), respectively , \(\forall i\), \eqref{eq:per-relay jamming constraint for DPS} can be rewritten as
\begin{align}\label{eq:hyperbolic per-relay jamming constraint for DPS}
z_i\le&c_{0,i}\left(1-\frac{y_i(c_{1,i} x_i+\sigma_{n_c}^2y_i)}{y_i-x_i}\right)
\left(1-\frac{x_i}{y_i}\right)\notag\\
\Leftrightarrow &\frac{z_i}{c_{0,i}}\le 1-\frac{x_i}{y_i}-(c_{1,i}x_i+\sigma_{n_c}^2y_i)\notag\\
\Leftrightarrow &\left(\sigma_{n_c}y_i\right)^2+\left(\sqrt{\left(1-\frac{z_i}{c_{0,i}}\right)\tfrac{1}{c_{1,i}}}\right)^2\notag\\
\le&\left(1-\frac{z_i}{c_{0,i}}-c_{1,sr,i}x_i\right)\left(y_i+\tfrac{1}{c_{1,i}}\right).
\end{align}
According to \eqref{eq:per-relay jamming constraint} and \eqref{eq:u1 and u2}, it is easily verified that \(1-\tfrac{z_i}{c_{0,sr,i}}-c_{1,sr,i}x_i>1-\rho_i\alpha_i-(1-\rho_i)\alpha_i\ge 0\). Hence, \eqref{eq:hyperbolic per-relay jamming constraint for DPS} is eligible for Lemma~\ref{lemma:hyperbolic to SOC}, which is reformulated into the SOC constraint:
\begin{align}\label{eq:SOC per-relay jamming constraint}
\left\|\begin{array}{c}2\sigma_{n_c}y_i\\
2\sqrt{\left(1-\tfrac{z_i}{c_{0,i}}\right)\tfrac{1}{c_{1,i}}}\\
\left(1-\tfrac{z_i}{c_{0,i}}-c_{1,i}x_i\right)-\left(y_i+\tfrac{1}{c_{1,i}}\right)
\end{array}\right\|\le \left(1-\tfrac{z_i}{c_{0,i}}-c_{1,i}x_i\right)+\left(y_i+\tfrac{1}{c_{1,i}}\right).
\end{align}
Similarly, \eqref{eq:constraint on rho} can be simplified as \(y_i(c_{1,i}x_i+\sigma_{n_c}^2y_i)\le y_i-x_i\), and after some manipulation, it is recast into a constraint of the restricted hyperbolic form as
\begin{equation}\label{eq:hyperbolic constraint on rho}
\left(\sigma_{n_c}y_i\right)^2+\left(\sqrt{\tfrac{1}{c_{1,i}}}\right)^2\le
\left(1-c_{1,i}x_i\right)\left(y_i+\tfrac{1}{c_{1,i}}\right).
\end{equation}
 \eqref{eq:hyperbolic constraint on rho} is thus, in line with Lemma~\ref{lemma:hyperbolic to SOC}, equivalent to an SOC constraint given by
\begin{align}\label{eq:SOC constraint on rho}
\left\|\begin{array}{c}2\sigma_{n_c}y_i\\
2\sqrt{\tfrac{1}{c_{1,i}}}\\
\left(1-c_{1,i}x_i\right)-\left(y_i+\tfrac{1}{c_{1,i}}\right)
\end{array}\right\|
\le \left(1-c_{1,i}x_i\right)+\left(y_i+\tfrac{1}{c_{1,i}}\right).
\end{align}
At last, \eqref{eq:constraint on alpha} is a linear constraint w.r.t.~\(x_i\) and \(y_i\) given by
\begin{equation}\label{eq:linear constraint on alpha}
  x_i-y_i\le 0,\; \forall i.
\end{equation}

\begin{figure*}
\begin{align}
   & \frac{\tau P_s{\rm trace}(\boldsymbol U_1\boldsymbol s_{sd}^\dag\boldsymbol s_{sd}^T)}{{\rm trace}(\overline{\boldsymbol S}\boldsymbol h_{rd}^\dag\boldsymbol h_{rd}^T)+{\rm trace}((\sigma_{n_a}^2\boldsymbol U_1+\sigma_{n_c}^2\boldsymbol U_2){\rm diag}(\boldsymbol c_{0}\circ\|\boldsymbol h_{rd}\|.^2))+\sigma_{n_d}^2}\label{eq:obj for (P2'.1-SDR)}\\
   & 1+\frac{P_s{\rm trace}(\boldsymbol U_1\boldsymbol s_{se,k}^\dag\boldsymbol s_{se,k}^T)}{{\rm trace}(\overline{\boldsymbol S}\boldsymbol h_{re,k}^\dag\boldsymbol h_{re,k}^T)+{\rm trace}((\sigma_{n_a}^2\boldsymbol U_1+\sigma_{n_c}^2\boldsymbol U_2){\rm diag}(\boldsymbol c_{0}\circ\|\boldsymbol h_{re,k}\|.^2))+\sigma_{n_e,k}^2}\le\frac{1}{\tau} \label{eq:SINR constraint for (P2'.1-SDR)}
  \end{align}
  \hrulefill
\end{figure*}

Note that \eqref{eq:per-relay jamming constraint for DPS}--\eqref{eq:constraint on rho} have so far been equivalently transformed into the SOC constraints \eqref{eq:SOC per-relay jamming constraint}, the linear constraints \eqref{eq:linear constraint on alpha}, as well as \eqref{eq:SOC constraint on rho}, the latter two of which are jointly convex w.r.t.~\(x_i\) and \(y_i\), \(\forall i\). However, \eqref{eq:SOC per-relay jamming constraint} is still not convex w.r.t.~\(z_i\), \(\forall i\), yet. To circumvent this, in the sequel we propose to solve problem \(\mathrm{(P2)}\) by alternating optimization. The upshot of the algorithm is that first we fix \(\boldsymbol S\) by \(\overline{\boldsymbol S}\) and thus \(z_i\) by \(\bar z_i={\rm trace}(\overline{\boldsymbol S}\boldsymbol E_i)\), \(\forall i\), and solve problem \(\mathrm{(P2^\prime)}\)\footnote{Note that we denote problem \(\mathrm{(P2)}\) (\(\mathrm{(P2.1)}\),\(\mathrm{(P2.2)}\)) with fixed \(\boldsymbol S\) as \(\mathrm{(P2^\prime)}\) (\(\mathrm{(P2^\prime.1)}\),\(\mathrm{(P2^\prime.2)}\)) in the sequel.} to find the optimal \(\{\alpha^\ast\}\), \(\{\rho^\ast\}\) and \(\{\measuredangle\beta_i\}\) via \(\mathrm{(P2^\prime.1)}\) and \(\mathrm{(P2^\prime.2)}\); then with \(\bar\alpha_i=\alpha^\ast_i\), \(\forall i\), we devise the optimal solution derived in Section~\ref{sec:optimal solutions for SPS} to obtain the optimal CJ covariance, viz \(\boldsymbol S^\ast\), and thus \(z_i^\ast={\rm trace}(\boldsymbol S^\ast\boldsymbol E_i)\), \(\forall i\); finally, by updating \(\overline{\boldsymbol S}=\boldsymbol S^\ast\) and \(\bar z_i=z_i^\ast\), \(\forall i\), problems \(\mathrm{(P2^\prime)}\) and \(\mathrm{(P1)}\) are  iteratively solved until they converge.

The remaining challenges lie in solving problem \(\mathrm{(P2^\prime.1)}\) now that \eqref{eq:SOC per-relay jamming constraint}, \eqref{eq:SOC constraint on rho} and \eqref{eq:linear constraint on alpha} are all made convex w.r.t.~their variables \(x_i\), \(y_i\), \(\forall i\). Similar to that for \(\mathrm{(P1.1)}\), we introduce \(\boldsymbol U_1=\boldsymbol u_1\boldsymbol u_1^H\) and \(\boldsymbol U_2=\boldsymbol u_2\boldsymbol u_2^H\) and exempt problem \(\mathrm{(P2^\prime.1)}\) from \({\rm rank}(\boldsymbol U_1)=1\) and \({\rm rank}(\boldsymbol U_2)=1\) as follows:
\begin{align*}
&~\mathrm{(P2^\prime.1\text{-}SDR)}:\\
&\left\{\begin{aligned}
\!\!\!\max_{\boldsymbol U_1,\boldsymbol U_2\succeq \boldsymbol 0, \{x_i\},\{y_i\}} \!\!\!&~H_2(\tau)\\
{\rm s.t.}&~\eqref{eq:SINR constraint for (P2'.1-SDR)}, \; \forall k, \ \eqref{eq:SOC per-relay jamming constraint}, \ \eqref{eq:SOC constraint on rho}, \ \eqref{eq:linear constraint on alpha},\\
{\rm trace}&(\boldsymbol U_1\boldsymbol E_i)=x_i, \, {\rm trace}(\boldsymbol U_2\boldsymbol E_i)=y_i,\; \forall i.
\end{aligned}\right.
\end{align*}
\(H_2(\tau)\) (c.f.~\eqref{eq:obj for (P2'.1-SDR)}) and \eqref{eq:SINR constraint for (P2'.1-SDR)} are given at the top of next page.
Recalling the procedure to deal with \(\mathrm{(P1.1\text{-}SDR)}\), we now apply Charnes-Cooper transformation to convert \(\mathrm{(P2^\prime.1\text{-}SDR)}\) into a convex problem, denoted  by \(\mathrm{(P2^\prime.1\text{-}SDP)}\), by replacing \(\boldsymbol U_1\) and \(\boldsymbol U_2\) with \(\hat{\boldsymbol U}_1/\xi\) and \(\hat{\boldsymbol U}_2/\xi\), respectively. The solution for \(\mathrm{(P2^\prime.1\text{-}SDP)}\) is tight and characterized by the following proposition.

\begin{proposition}
We have the following results:
\begin{enumerate}[1)]
  \item The optimal solution to \(\mathrm{(P2^\prime.1\text{-}SDP)}\) satisfies \({\rm rank}(\hat{\boldsymbol U}_1^\ast)=1\) such that \(\hat{\mv U}_1^\ast=\hat{\mv u}_1^\ast\hat{\mv u}_1^{\ast H}\);
 \item \(\hat{\boldsymbol u}_1^\ast\) is given by
 \begin{equation}\label{eq:optimal u1}
 \hat{\boldsymbol u}_1^\ast=
 \sqrt{\frac{\begin{array}{r}
 \tau-\xi^\ast\sigma_{n_d}^2-\sigma_{n_c}^2{\rm trace}(\hat{\boldsymbol U}_2^\ast\boldsymbol C_{rd} )\\
 -\xi^\ast{\rm trace}(\overline{\boldsymbol S}\boldsymbol h_{rd}^\dag\boldsymbol h_{rd}^T)
 \end{array}}{\sigma_{n_a}^2{\rm trace}(\hat{\boldsymbol u}_1\hat{\boldsymbol u}_1^H\boldsymbol C_{rd})}}\hat{\boldsymbol u}_1,
\end{equation}
where \(\hat{\boldsymbol u}_1=(\boldsymbol\Xi^\prime+\sum_{k=1}^K\theta_k^\ast P_s\boldsymbol s_{se,k}^\dag\boldsymbol s_{se,k}^T)^{-1}\boldsymbol s_{sd}^\dag\), \(\boldsymbol C_{rd}={\rm diag}(\boldsymbol c_0\circ\|\boldsymbol h_{rd}\|.^2)\), and \(\boldsymbol\Xi^\prime\) is given in Appendix~\ref{appendix:proof of rank-one solution with CJ for DPS};
\item \(\hat{\boldsymbol U}_2^\ast\), of which the diagonal entries compose a vector denoted by \(\hat{\boldsymbol u}_2^\ast\), can be reconstructed by \(\hat{\boldsymbol u}_2^\ast\hat{\boldsymbol u}_2^{\ast H}\) with rank one.
\end{enumerate}
\label{prop:rank-one solution with CJ for DPS}
\end{proposition}

\begin{IEEEproof}
See Appendix~\ref{appendix:proof of rank-one solution with CJ for DPS}.
\end{IEEEproof}

The \(\alpha_i^\ast\)'s and \(\rho_i\)'s are thus attained according to \eqref{eq:alpha and rho} via EVD of \(\boldsymbol U_1^\ast\) and \(\boldsymbol U_2^\ast\). The proposed algorithm for solving \(\mathrm{(P2)}\) is presented in Table~\ref{table:Algorithm I}.
\begin{table}[!ht]
\begin{center}
\caption{{\rm Algorithm for Solving \(\mathrm{(P2)}\)} }\label{table:Algorithm I}
\vspace{-0.5em}
 \hrule
\vspace{0.75em}
\begin{algorithmic}[1]
\REQUIRE \(\boldsymbol S^\ast\); \(r_{\rm SPS}^\ast\) that denotes the optimum value for \(\mathrm{(P1)}\) given \(\bar\alpha_i=.5\), \(\forall i\)
\STATE \(ii\leftarrow 0\), \(r^{(ii)}_{\rm sec}\leftarrow r_{\rm SPS}^\ast\)
\REPEAT
\STATE\(ii\leftarrow ii+1\)
\STATE \(\bar{\boldsymbol S} \leftarrow\boldsymbol S^\ast\) , \(\bar z_i\leftarrow{\rm trace}(\boldsymbol S^\ast\boldsymbol E_i)\),  \(\forall i\), and solve \(\mathrm{(P2^\prime)}\):
\STATE \(kk\leftarrow 0\), \(r_{\rm DPS}^{(0)}\leftarrow 10^{-6}\), \(r_{\rm DPS}^{(1)}\leftarrow 10\), \(l\leftarrow\tau_{\min,2}\), \(u\leftarrow1\)
\WHILE{\(\vert r_{\rm DPS}^{(kk+1)}-r_{\rm DPS}^{(kk)}\vert/r_{\rm DPS}^{(kk)}>\epsilon_b\)}
\STATE \(kk\leftarrow kk+1\), \(\tau\leftarrow\frac{l+u}{2}\)
\STATE solve \(\mathrm{(P2^\prime.1)}\) and \RETURN \(H_2(\tau)\)
\STATE \(r_{\rm DPS}^{(kk+1)}\leftarrow\frac{1}{2}\log_2(\tau+H_2(\tau))\)
\STATE \(r_{\rm temp}\leftarrow\frac{1}{2}\log_2(\tilde\tau+H_2(\tilde\tau))\), where \(\tilde\tau\leftarrow\max(\tau-\Delta\tau,\tau_{\min,1})\) and \(\Delta\tau>0\) denotes an arbitrary small value.
\IF{\(r_{\rm DPS}^{(kk+1)}\le r_{\rm temp}\)}
\STATE \(u\leftarrow\tau\)
 \ELSE
\STATE \(l\leftarrow\tau\)
 \ENDIF
\ENDWHILE
\RETURN \(\mv U_1^\ast\), \(U_2^\ast\), and obtain \(\{\alpha_i^\ast\}\) according to \eqref{eq:alpha and rho}
\STATE \(\bar\alpha_i\leftarrow\alpha_i^\ast\), \(\forall i\), and solve \(\mathrm{(P1)}\) via \(\mathrm{(P1.1)}\) and \(\mathrm{(P1.2)}\)
\RETURN \(\mv X_1^\ast\), \(\mv S^\ast\), and obtain \(\{\rho_i^\ast\}\) and \(\{\measuredangle\beta_i^\ast\}\) according to \(w_{1,i}^\ast=\sqrt{1-\rho_i^\ast}e^{j\measuredangle\beta_i^\ast}\), \(\forall i\)
\STATE Update \(r_{\rm sec}^{(ii)}\) according to \eqref{eq:secrecy rate region}
\UNTIL{\(r^{(ii)}_{\rm sec}-r^{(ii-1)}_{\rm sec}\le\epsilon_0\)}
\ENSURE \(\{\alpha_i^\ast\}\), \(\{\rho_i^\ast\}\), \(\{\measuredangle\beta_i^\ast\}\), and \(\boldsymbol S^\ast\)
\end{algorithmic}
\vspace{0.75em}
\hrule
\end{center}
\vspace{-1.0em}
\end{table}

\section{Distributed Algorithms}\label{sec:distributed}
In this section, we investigate heuristic algorithms to solve problems \(\mathrm{(P1)}\) and \(\mathrm{(P2)}\) in a completely distributed fashion. Note that different from the paradigm of {\em distributed optimization} that allows for certain amount of information exchange based on which iterative algorithms are developed to gradually improve the system performance, we herein assume that each individual relay can only make decision based on its local CSIs, namely, \(h_{sr_i}\), \(h_{r_id}\), \(h_{r_ie}\), \(\forall i\), and there is no extra means of information acquisition for ease of implementation. The purpose for such an algorithm is twofold: on one hand, we aim to answer the question that in the least favourable situation, namely, no coordination over the relays, how to improve the achievable secrecy rate of the system? On the other hand, it provides a lower-bound for the centralized schemes proposed in Section~\ref{sec:centralized}, which sheds light upon the trade-off achievable between secrecy performance and complexity.

Besides, we emphasize the jamming scheme that is different from the CJ in the centralized schemes. Unlike the CJ signal coordinately transmitted by all relays, in the distributed implementation, each relay is only able to generate its AN locally, i.e.,  \(\boldsymbol x_{r2}=[\sqrt{\sigma_1}s_1^\prime, \ldots, \sqrt{\sigma_N}s_N^\prime]^T\), in which \(s_i\)'s are i.i.d.~AN beams, denoted by \(s_i^\prime\sim\mathcal{CN}(0,1)\). This type of CJ is known to be uncoordinated with the covariance matrix given by \(\boldsymbol S={\rm diag}([\sigma_1,\ldots,\sigma_N])\). In this section, we assume that each relay consumes all of its remaining power from AF for AN, i.e., \(\sigma_i=\eta\rho_i\alpha_i P_s\vert h_{sr_i}\vert^2\), \(\forall i\in\mathcal{N}\) (c.f.~\eqref{eq:per-relay jamming constraint}). Hence, the AN design solely depends on \(\alpha_i\)'s and/or  \(\rho_i\)'s.

\subsection{Distributed Algorithm for SPS}\label{sec:distributed alg. for SPS}
First, we propose a heuristic scheme for the \(i\)th AF relay to decide on \(\rho_i\), \(\forall i\), which is given by
\begin{align}\label{eq:rho for distributed alg}
  \rho_i=\delta\left(1-\frac{\vert h_{r_id}\vert^2}{\max\limits_{k\in\mathcal{K}}\vert h_{r_ie,k}\vert^2}\right)^+,
\end{align}
where \(\delta\in(0,1)\) is a constant controlling the relay's level of jamming. For example, a larger \(\delta\) indicates that each relay prefers to splitting  a larger portion of power for jamming and vice versa. The intuition behind \eqref{eq:rho for distributed alg} is that if the \(i\)th relay observes that \(\vert h_{r_id}\vert^2\ge\max\limits_{k\mathcal{K}}\vert h_{r_ie,k}\vert^2\), which means that a nonnegative secrecy rate is achievable even if there is only itself in the system, it will shut down the AN; otherwise, it will split up to \(\delta\) portion of the harvested power for jamming. In an extreme case of \(\vert h_{r_id}\vert^2\ll\max\limits_{k\mathcal{K}}\vert h_{r_ie,k}\vert^2\), probably when an Eve is located within the very proximity of this relay, it allocates the maximum permissible portion of  power, i.e., \(\delta\),  for AN.

Next, since an individual relay cannot evaluate the secrecy performance of the whole system, \(\measuredangle\beta_i\)'s are simply chosen to be the optimum for the multi-AF relaying without security considerations, i.e., \(\measuredangle\beta_i=-\measuredangle h_{r_id}-\measuredangle h_{sr_i}\), \(\forall i\).

\subsection{Distributed Algorithm for DPS}\label{sec:distributed alg. for DPS}
Following the same designs for \(\rho_i\)'s and \(\measuredangle\beta_i\)'s in Section~\ref{sec:distributed alg. for SPS}, the remaining task for WEH-enabled relays operating with DPS is to set  proper values for \(\alpha_i\)'s. We choose \(\alpha_i\)'s that maximize the ``hypothetical SINR''. This ``hypothetical'' SINR may not be the actual SINR for the destination, but just a criterion calculated based on the ``hypothetical'' received signal seen by the \(i\)th relay, given by
\begin{align}
\tilde y_{d_i}=h_{r_id}\beta_i\sqrt{1-\alpha_i}\sqrt{P_s}h_{sr_i}s+h_{r_id}\beta_i\sqrt{1-\alpha_i}n_{a,i}
+h_{r_id}\beta_in_{c,i}+h_{r_id}\sqrt{\sigma_i}s_i^\prime+n_d, \; \forall i.\label{eq:hypo received signal at the Rx}
\end{align}
The corresponding SINR is thus expressed as
\begin{equation}\label{eq:hypo SINR at the Rx}
{\rm SINR}_{\tilde y_{d_i}}=
\frac{\eta (1-\rho_i)P_s\vert h_{sr_i}\vert^2}{
\eta\sigma_{n_a}^2+\frac{\eta\sigma_{n_c}^2}{1-\alpha_i}
+\frac{\gamma_{i}(P_s\vert h_{sr_i}\vert^2+\sigma_{n_a}^2)}{\alpha_i}+\frac{\gamma_{i}\sigma_{n_c}^2}{\alpha_i(1-\alpha_i)}+\eta\rho_iP_s\vert h_{sr_i}\vert^2},
\end{equation}
where \(\gamma_{i}=\tfrac{\sigma_{n_d}^2}{P_s\vert h_{sr_i}\vert^2\vert h_{r_id}\vert^2}\). Consequently, the maximization of \eqref{eq:hypo SINR at the Rx} w.r.t.~\(\alpha_i\), \(\forall i\), is formulated as
\begin{align*}
&\mathrm{(P2\text{-}distr.)}:\\
& \left\{\begin{aligned}
\min_{\alpha_i} &~\frac{\eta\sigma_{n_c}^2}{1-\alpha_i}
    +\frac{\gamma_{i}(P_s\vert h_{sr_i}\vert^2+\sigma_{n_a}^2)}{\alpha_i}+
    \frac{\gamma_{i}\sigma_{n_c}^2}{\alpha_i(1-\alpha_i)}\\
{\rm s.t.} &~0\le\alpha_i\le1.
\end{aligned}\right.
\end{align*}

\begin{proposition}
The optimal \(\alpha_i\), \(\forall i\), to \(\mathrm{(P2\text{-}distr.)}\) is
\begin{equation}\label{eq:optimal alpha to sub1}
\alpha_i^\ast=\frac{1}{1+\sqrt{\frac{(\eta+\gamma_{i})\sigma_{n_c}^2}{\gamma_{i}(P_s\vert h_{sr_i}\vert^2+\sigma_{n_a}^2+\sigma_{n_c}^2)}}}.
\end{equation}\label{prop:optimal alpha to distr}
\end{proposition}

\begin{IEEEproof}
It is easy to verify that problem \(\mathrm{(P2\text{-}distr.)}\) is convex and the minimum solution of its objective function derived from the first-order derivative happens to fall within the feasible region of \(\alpha_i\), which is seen in \eqref{eq:optimal alpha to sub1}.
\end{IEEEproof}

With \(\rho_i\)'s, \(\measuredangle\beta_i\)'s and \(\alpha_i\)'s set, each AF relay is then able to decide its relay weight and AN transmission.

\section{Numerical Results}\label{sec:Numerical Results}
In this section we compare our proposed schemes operating with SPS or DPS with some benchmarks. In the centralized case, the optimal solution for SPS in Section~\ref{sec:optimal solutions for SPS} is denoted by {\em CJ-SPS}, while Algorithm~\ref{table:Algorithm I} in Section~\ref{sec:proposed solutions for DPS} is denoted by {\em CJ-DPS}. The distributed schemes in Section~\ref{sec:distributed alg. for SPS} and Section~\ref{sec:distributed alg. for DPS} are referred to as {\em Distributed-SPS} and {\em Distributed-DPS}, respectively. To demonstrate the effectiveness of our AN-aided secure multi-AF relay beamforming algorithms, we also provide three benchmark schemes: {\em NoCJ-SPS}, {\em NoCJ-DPS} and {\em Random PS}. For {\em NoCJ-SPS}, we solve problem \(\mathrm{(P1)}\) by replacing \(\boldsymbol S\) with \(\mv 0\). Similarly, for {\em NoCJ-DPS}, we initialize \(\overline{\boldsymbol S}=\mv 0\) and quit the loop in Algorithm~\ref{table:Algorithm I} after the very first time of solving problem \(\mathrm{(P2^\prime)}\). {\em Random PS}, on the other hand, picks up i.i.d.~\(\alpha_i\) and \(\rho_i\) uniformly generated over \([0,1]\), respectively, and co-phases \(\measuredangle\beta_i=-\measuredangle h_{sr_i}-\measuredangle h_{r_id}\), \(\forall i\).

Consider that $N$ WEH-enabled AF relays and $K$ eavesdroppers are located within a circular area of radius $R$. Specifically, we assume that their respective radius and radian are drawn from uniform distributions over the interval \([0,R]\) and \([0,2\pi)\), respectively. We also assume that the channel models consist of both large-scale path loss and small-scale multi-path fading.
The unified path loss model is given by
 \begin{align}\label{eq:path loss model}
   L=A_0\left(\frac{d}{d_0}\right)^{-\alpha},
 \end{align}
where \(A_0=10^{-3}\), \(d\) denotes the relevant distance, $d_0=1$m is a reference distance, and \(\alpha\) is the path loss exponent set to be \(2.5\). \(h_{sr_i}\), \(h_{r_id}\), and \( h_{r_ie,k}\), \(\forall i\in\mathcal{N}\), \(\forall k\in\mathcal{K}\), are generated from independent Rayleigh fading with zero mean and variance specified by \eqref{eq:path loss model}.

The simulation parameters are set as follows unless otherwise specified: the radius defining the range is $R=5$m; the transmit power at the source is $P_s=40$dBm; the noise variances are set as \(\sigma_{n_a}^2=-50\)dBm, \(\sigma_{n_c}^2=-80\)dBm, \(\sigma_{n_d}^2=\sigma_{n_a}^2+\sigma_{n_c}^2\), and \(\sigma_{n_e,k}^2=\sigma_{n_d}^2\), \(\forall k\); the EH efficiency is set to $\eta=50\%$. Also, numerical results were averages over $500$ independent channel realizations.

\subsection{Secrecy Performance by Centralized Approach}\label{sec:Performance by Centralized Approach}
\begin{figure}[htp]
  \begin{center}
   \includegraphics[width=3.0in]{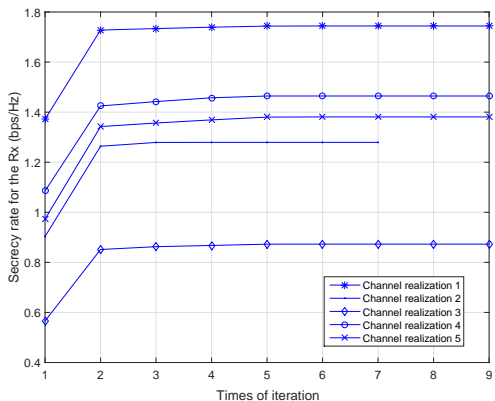}
  \end{center}
 \caption{The secrecy rate by {\em CJ-DPS} versus the number of iterations for the alternating optimization in Table I with $P_s=40$dBm, $N=10$, and $K=5$.}\label{fig:alternating sec vs Iter}
\end{figure}

Here, we evaluate the performance of the proposed centralized designs in Section~\ref{sec:centralized}. The efficiency of the alternating optimization that iteratively attains numerical solution to \(\mathrm{(P2)}\) is studied in Fig.~\ref{fig:alternating sec vs Iter}, which shows the increment of the achievable secrecy rate after each round of the iteration. The most rapid increase is observed after the first iteration, which illustrates that the optimization of the PS ratios, \(\alpha_i\)'s, accounts for the main factor for the secrecy rate performance gains over a SPS scheme that sets \(\{\alpha_i=0.5\}\). It is seen that the alternating algorithm converges within the relative tolerance set to be \(10^{-3}\),  after an average of $5$--$6$ iterations for several channel realizations, which appears reasonable in terms of complexity.

\begin{figure}[htp]
\centering
\subfigure[{\(\max\limits_k r_{\rm S,E,k}\) versus the number of relays.}]{\label{fig:subfig:Eve's rate vs N}\includegraphics[width=3.0in]{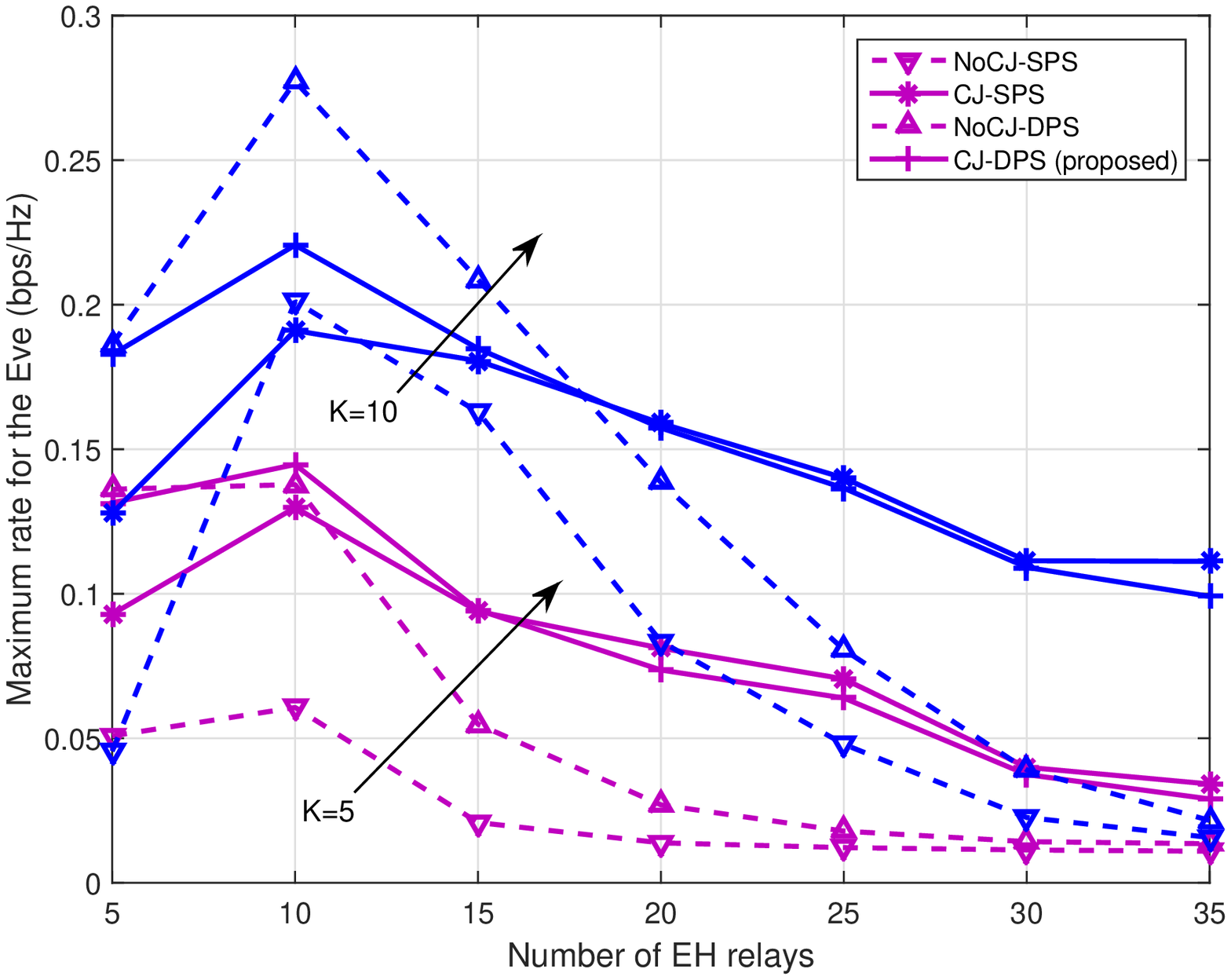}}
\subfigure[{The achievable secrecy rate vs the number of AF relays.}]
{\label{fig:subfig:sec vs N}\includegraphics[width=3.0in]{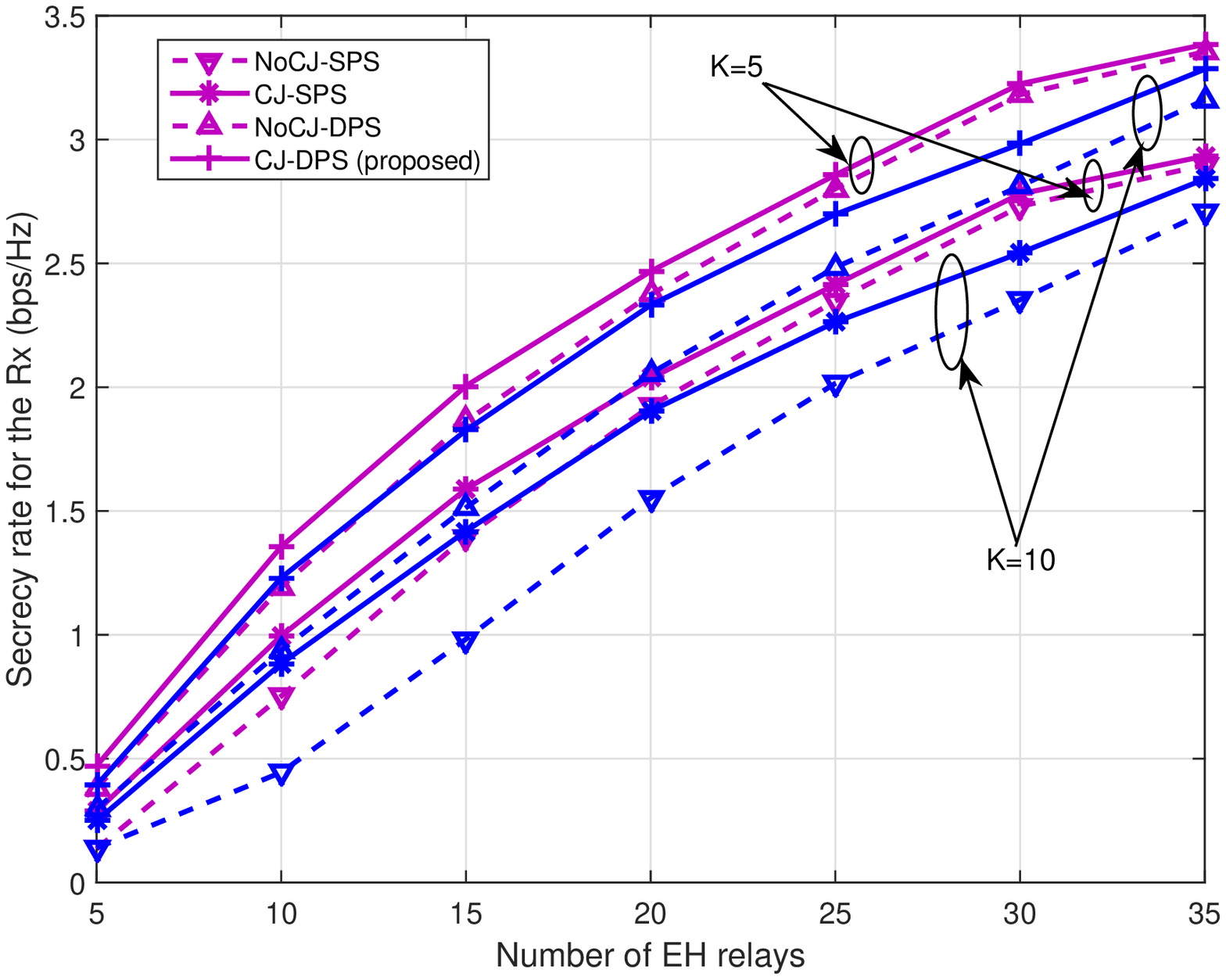}}
\caption{Comparison of different schemes with $P_s=10$dB for $K=5$ and $K=10$, respectively.}\label{fig:centralized compare vs N}
\end{figure}

Fig.~\ref{fig:centralized compare vs N} shows the achievable secrecy rate for the legitimate Rx with the increase in the number of AF relays by different schemes. As we can see, the secure multi-AF relaying schemes assisted by the transmission of  AN outperforms that without AN for both SPS and DPS. In addition, with the increase in $N$, the role of CJ gradually reduces for both schemes of SPS and DPS. This is because as $N$ gets larger, the optimal designs tend to suppress the interception at the most capable eavesdropper more effectively with $N$ DoF, enforcing the numerator of \(\max_{k\in\mathcal{K}}{\rm SINR}_{\rm S,E,k}\) to a relatively low level, which can also be observed from \(\max_{k\in\mathcal{K}}r_{S,E,k}\) in Fig.~\ref{fig:subfig:Eve's rate vs N}, and therefore the optimal amount of power allocated to AN beams inclines to be little; otherwise the jamming yielded will be detrimental to the reception at the legitimate Rx. Besides, given the same number of AF relays, the secrecy performance gains brought by the proposed schemes with CJ are more prominent in the presence of more eavesdroppers, since it is hard to reduce all the eavesdroppers' channel capacity without resorting to CJ properly.

\begin{figure}[htp]
  \begin{center}
   \includegraphics[width=3.0in]{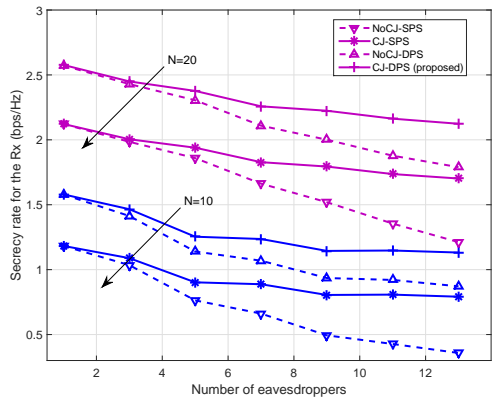}
  \end{center}
 \caption{The secrecy rate versus the number of eavesdroppers with $P_s=10$dB for $N=10$ and $N=20$, respectively.}\label{fig:centralized compare sec vs Eves}
\end{figure}

Fig.~\ref{fig:centralized compare sec vs Eves} shows the achievable secrecy rate for the legitimate Rx versus the number of eavesdroppers by different schemes. First, similar to the results shown in Fig.~\ref{fig:centralized compare vs N}, the proposed AN-aided multi-AF relaying designs operating with DPS-enabled relays, viz {\em CJ-DPS}, perform best among all the schemes. Secondly, as $K$ goes up, the AN-aided schemes, {\em CJ-DPS} and {\em CJ-SPS}, allow the secrecy rate to drop slowly, in other words, more robust against multiple eavesdroppers, while the secrecy rate of their NoCJ counterparts almost goes down linearly with $K$. Furthermore, with $K$ increasing, for example, more than $10$, the increase in the number of relays, from $N=10$ to $20$, cannot replace the role of CJ as shown in Fig.~\ref{fig:centralized compare vs N} and \ref{fig:centralized compare sec vs Eves}, since in the presence of many eavesdroppers, more relays may also result in improved eavesdroppers' decoding ability w/o the assistance of CJ. It is also noteworthy that with $K=1$, there is little use of CJ by the centralized schemes, which was also observed in \cite{Li2013}.

\begin{figure}[htp]
  \begin{center}
   \includegraphics[width=3.0in]{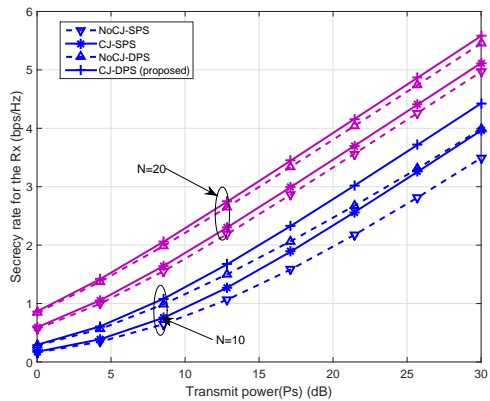}
  \end{center}
 \caption{The secrecy rate versus the transmit power with $K=5$ for $N=10$ and $N=20$, respectively.}\label{fig:centralized compare sec vs Ps}
\end{figure}

Fig.~\ref{fig:centralized compare sec vs Ps} provides simulation results of different schemes by varying the source transmit power. It is seen that with more power available at the source, the advantage of CJ is more pronounced, since given other variables fixed, larger $P_s$ indicates larger feasible regions for \(\mathrm{(P1)}\) and \(\mathrm{(P2)}\). Furthermore, as similarly seen in Fig.~\ref{fig:centralized compare vs N}, with a mild number of eavesdroppers ($K=5$), subject to the same $P_s$, a large number of cooperative relays enables more DoF in designing the optimal \(\alpha_i\)'s and \(\measuredangle\beta_i\)'s, which alleviates the dependence on AN beams to combat Eves.

\subsection{Secrecy Performance by Distributed Algorithms}\label{sec:Performance by Distributed Implementation}
Here, we study the performance of the distributed schemes, namely, {\em Distributed-SPS} and {\em Distributed-DPS} in Section~\ref{sec:distributed}. As mentioned earlier, these heuristics are provided as benchmarks to demonstrate what can be done under the extreme ``no-coordination'' circumstance, in comparison with {\em Random PS}. Note that any other distributed schemes with certain level of cooperation among relays are supposed to increase the secrecy performance up to the proposed centralized algorithms, namely, {\em CJ-DPS} and {\em CJ-SPS}, at the expense of extra computational complexity and system overhead.

\begin{figure}[htp]
  \begin{center}
   \includegraphics[width=3.0in]{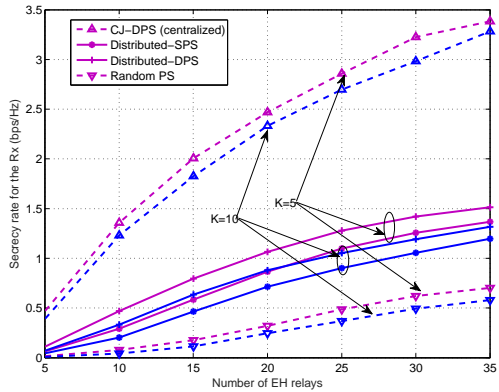}
  \end{center}
 \caption{The secrecy rate versus the number of AF relays by distributed algorithms with $P_s=10$dB and $K=5$.}\label{fig:distributed compare sec vs N}
\end{figure}

Fig.~\ref{fig:distributed compare sec vs N} provides the results for the achievable secrecy rate of various schemes versus the number of relays. {\em Distributed-SPS} and {\em Distributed-DPS}, are observed to be outperformed by their centralized counterparts though, they are considerably superior to {\em Random PS}. It is also seen that the performance gap between the centralized and distributed approaches is enlarged as $N$ increases, which is expected, since larger $N$ yields more DoF for cooperation that is exclusively beneficial for the centralized schemes. Furthermore, compared with the centralized schemes, the distributed ones are more vulnerable to the increase in the eavesdroppers' number.

\begin{figure}[htp]
  \begin{center}
   \includegraphics[width=3.0in]{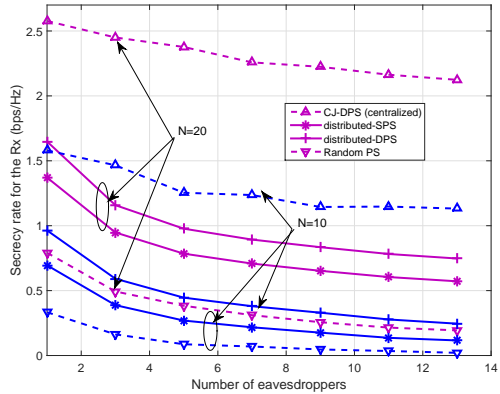}
  \end{center}
 \caption{The secrecy rate versus the number of eavesdroppers by distributed algorithms with $P_s=10$dB and $N$=8.}\label{fig:distributed compare sec vs Eves}
\end{figure}

In Fig.~\ref{fig:distributed compare sec vs Eves}, we investigate the relationship between the secrecy rate performance and the number of eavesdroppers by different methods. As can be observed, compared with the centralized schemes, the secrecy rates achieved by {\em Distributed-SPS} and {\em Distributed-DPS} both reduce more drastically with the increase in $K$ due to the lack of effective cooperation. Also, the advantage of DPS over SPS for the distributed schemes is compromised since \(\alpha_i\)'s are not jointly designed with other parameters. At last, a similar observation has been made as that for Fig.~\ref{fig:distributed compare sec vs N}, that is, larger $N$ yields more visible performance gap between the centralized and distributed approaches.

\begin{figure}[htp]
  \begin{center}
   \includegraphics[width=3.0in]{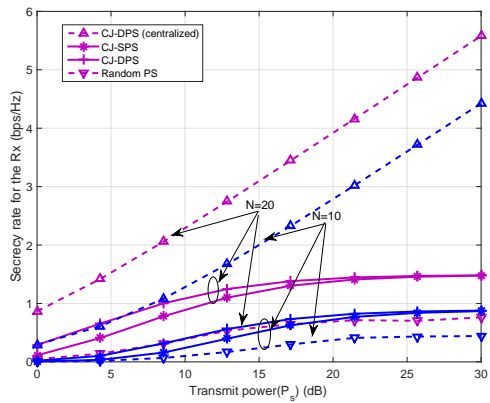}
  \end{center}
 \caption{The secrecy rate versus transmit power by distributed algorithms with $N=10$ and $K=5$.}\label{fig:distributed compare sec vs Ps}
\end{figure}

In Fig.~\ref{fig:distributed compare sec vs Ps}, we examine the effect of increasing the transmit power at the source on the secrecy performance of different schemes under the same settings as those in Fig.~\ref{fig:centralized compare sec vs Ps}. Among all the presented designs, {\em CJ-DPS} still achieves the best secrecy rate as observed in other examples. Also, the fact that larger $N$ benefits more from cooperative designs is corroborated again due to the same reason as that for Fig.~\ref{fig:distributed compare sec vs N} and \ref{fig:distributed compare sec vs Eves}. Furthermore, the secrecy rate of {\em Distributed-SPS} or {\em Distributed-DPS} is quickly saturated when $P_s>20$dB while that for their centralized counterparts still rises at fast speed.

\section{conclusion}\label{sec:conclusion}
This paper studied secure communications using multiple single-antenna WEH-enabled AF relays assisted by AN via CJ for a SWIPT network with multiple single-antenna eavesdroppers. Using PS at the relays, the achievable secrecy rates for the relay wiretap channel were maximized by jointly optimizing the CB and the CJ covariance matrix along with the PS ratios for relays operating with, respectively, SPS and DPS. For DPS, Reformulating the constraints into restricted hyperbolic forms enabled us to develop convex optimization-based solutions. Further, we proposed an information-exchange-free distributed algorithm that outperforms the random decision.

\begin{appendix}
\subsection{Proof of Lemma~\ref{lemma:concavity of H_1(tau)}}\label{appendix:proof of concavity of H_1(tau)}
As the optimal solution to \(\mathrm{(P1.1\text{-}SDP)}\) is proved to be optimal to problem \(\mathrm{(P1.1)}\), the optimum value for \(\mathrm{(P1.1\text{-}SDP)}\) is \(\tau f_1(\tau)=H_1(\tau)\). Hence, we identify the property of \(H_1(\tau)\) by investigating \(\mathrm{(P1.1\text{-}SDP)}\), with its Lagrangian given by
\begin{align}
\mathcal{L}(\chi )=&
{\rm trace}\left(\left(\begin{array}{r}
P_s\tilde{\boldsymbol h}_{sd}^\dag\tilde{\boldsymbol h}_{sd}^T-\lambda\boldsymbol D_{\hat{sd}}\\
+\sum_{k=1}^K\theta_k\left(\frac{1}{\tau}-1\right)\boldsymbol D_{\hat{se},k}\\
-\sum_{k=1}^K\theta_k P_s\tilde{\boldsymbol h}_{se,k}^\dag\tilde{\boldsymbol h}_{se,k}^T\\
-\boldsymbol W_0^{\tfrac{1}{2}}\boldsymbol U\boldsymbol W_0^{\tfrac{1}{2}}+\boldsymbol Y_1
\end{array}\right)\hat{\boldsymbol X}_1\right)
+{\rm trace}\left(\left(\begin{array}{r}
-\lambda\boldsymbol h_{rd}^\dag\boldsymbol h_{rd}^T\\
+\sum_{k=1}^K\theta_k\left(\frac{1}{\tau}-1\right)\boldsymbol h_{re,k}^\dag\boldsymbol h_{re,k}^T\\
-\boldsymbol U+\boldsymbol Y_2
\end{array}\right)\hat{\boldsymbol S} \right)\notag\\
&+\left(\begin{array}{r}
-\lambda\sigma_{n_d}^2+\sum_{k=1}^K\theta_k\left(\frac{1}{\tau}-1\right)\sigma_{n_e,k}^2\\
+{\rm trace}\left(\boldsymbol W_0^{\frac{1}{2}}\boldsymbol U\boldsymbol W_0^{\frac{1}{2}}\right)+\zeta
\end{array}\right)\xi+\lambda\tau, \label{eq:Lagrangian of (P1.1-SDP)}
\end{align}
where \(\chi\) denotes a tuple consisting of all the primal and dual variables. Specifically, \(\boldsymbol Y_1\), \(\boldsymbol Y_2\) and \(\lambda\) are Lagrangian multipliers associated with \(\hat{\boldsymbol X}_1\), \(\hat{\boldsymbol S}\) and the first constraint of \(\mathrm{(P1.1\text{-}SDP)}\), respectively; \(\{\theta_k\}\) are the dual variables associated with the SINR constraint for the \(k\)th Eve, respectively; \(\boldsymbol U={\rm diag}([u_i]_{i=1}^N)\) with each diagonal entry \(u_i\) denoting the dual variable associated with the per-relay power constraint; \(\zeta\) is the Lagrangian multiplier associated with \(\xi\ge 0\). In addition, \(\boldsymbol W_0={\rm diag}([\eta\bar\alpha_iP_s|h_{sr_i}|^2]_{i=1}^N)\). The Karush-Kuhn-Tucker (KKT) conditions for \eqref{eq:Lagrangian of (P1.1-SDP)} are partially listed as follows:
\begin{subequations}\label{eq:KKT conditions for (P1.1-SDP)}
\begin{align}
&\boldsymbol Y_1 =-P_s\tilde{\boldsymbol h}_{sd}^\dag\tilde{\boldsymbol h}_{sd}^T+\lambda\boldsymbol D_{\hat{sd}}-\sum_{k=1}^K\theta_k\left(\frac{1}{\tau}-1\right)\boldsymbol D_{\hat{se},k}+\sum_{k=1}^K\theta_kP_s\tilde{\boldsymbol h}_{se,k}^\dag\tilde{\boldsymbol h}_{se,k}^T+\boldsymbol W_0^{\frac{1}{2}}\boldsymbol U\boldsymbol W_0^{\frac{1}{2}},\label{eq:Y1 for SPS}\\
&\boldsymbol Y_2 =\lambda\boldsymbol h_{rd}^\dag\boldsymbol h_{rd}^T-\sum_{k=1}^K\theta_k\left(\frac{1}{\tau}-1\right)\boldsymbol h_{re,k}^\dag\boldsymbol h_{re,k}^T+\boldsymbol U,\label{eq:Y2 for SPS}\\
&\zeta=\lambda\sigma_{n_d}^2-\sum_{k=1}^K\theta_k\left(\frac{1}{\tau}-1\right)\sigma_{n_e,k}^2-{\rm trace}(\boldsymbol W_0^{\frac{1}{2}}\boldsymbol U\boldsymbol W_0^{\frac{1}{2}}).\label{eq:zeta for SPS}
\end{align}
\end{subequations}

The associated dual problem is accordingly given by
\begin{align*}
&\mathrm{(P1.1\text{-}SDP}\mathrm{\text{-}dual)}:\\
&\left\{\begin{aligned}
\min_{\lambda,\{\theta_k\},\{u_i\}}&~\lambda\tau\\
{\rm s.t.}&~\theta_k\ge 0, \; \forall k, \ u_i\ge 0, \; \forall i, \\
&~\lambda\ge 0, \ \zeta\ge 0, \ \boldsymbol Y_1\succeq\boldsymbol 0, \ \boldsymbol Y_2\succeq\boldsymbol 0,
\end{aligned}\right.
\end{align*}
where \(\boldsymbol Y_1\), \(\boldsymbol Y_2\), and \(\zeta\) are given by \eqref{eq:Y1 for SPS}--\eqref{eq:zeta for SPS}, respectively. Since it is easily verified that \(\mathrm{(P1.1\text{-}SDP)}\) satisfies the Slater condition, the strong duality holds \cite{Boyd2004}. This implies that the dual optimum value in \(\mathrm{(P1.1\text{-}SDP\text{-}dual)}\) is \(H_1(\tau)\), which turns out to be a point-wise minimum of a family of affine functions and thus concave over \(\tau\in[\tau_{\min,1},1]\) \cite[pp.~80]{Boyd2004}.

\subsection{Proof of Proposition~\ref{prop:rank-one solution with CJ for SPS}}\label{appendix:proof of rank-one solution with CJ for SPS}
The KKT conditions for \eqref{eq:Lagrangian of (P1.1-SDP)} yield the following complementary slackness:
\begin{subequations}\label{eq:CS for (P1.1-SDP)}
\begin{align}
&\boldsymbol Y_1\hat{\boldsymbol X}_1^\ast =\boldsymbol 0,\label{eq:CS wrt hat X1}\\
&\boldsymbol Y_2\hat{\boldsymbol S}^\ast =\boldsymbol 0.\label{eq:CS wrt hat S}
\end{align}
\end{subequations}
Pre- and post-multiplying \(\boldsymbol W_0^{\frac{1}{2}}\) with the left hand side (LHS) and right hand side (RHS) of \eqref{eq:Y2 for SPS}, respectively, and substituting \(\boldsymbol W_0^{\frac{1}{2}}\boldsymbol U^\ast\boldsymbol W_0^{\frac{1}{2}}\) into \eqref{eq:Y1 for SPS}, \(\boldsymbol Y_1^\ast\) can be rewritten as
\begin{align}\label{eq:Y1 intermediate1}
  \boldsymbol Y_1 =& -P_s\tilde{\boldsymbol h}_{sd}^\dag\tilde{\boldsymbol h}_{sd}^T+\lambda\boldsymbol D_{\hat{sd}}-\sum_{k=1}^K\theta_k\left(\frac{1}{\tau}-1\right)\boldsymbol D_{\hat{se},k}
-\lambda\boldsymbol W_0^{\frac{1}{2}}\boldsymbol h_{rd}^\dag\boldsymbol h_{rd}^T\boldsymbol W_0^{\frac{1}{2}}+\sum_{k=1}^K\theta_kP_s\tilde{\boldsymbol h}_{se,k}^\dag\tilde{\boldsymbol h}_{se,k}^T\notag\\
&+\boldsymbol W_0^{\frac{1}{2}}\boldsymbol Y_2\boldsymbol W_0^{\frac{1}{2}}+\sum_{k=1}^K\theta_k\left(\frac{1}{\tau}-1\right)\boldsymbol W_0^{\frac{1}{2}}\boldsymbol h_{re,k}^\dag\boldsymbol h_{re,k}^T\boldsymbol W_0^{\frac{1}{2}}.
\end{align}
Introducing the notation of \([\cdot]_{\rm offd}\) to represent a square matrix with its diagonal entries removed, it follows from~\eqref{eq:Y2 for SPS} that
\begin{equation}\label{eq:off-diagonal cancellation}
\left[\begin{array}{r}
\boldsymbol W_0^{\frac{1}{2}}\boldsymbol Y_2\boldsymbol W_0^{\frac{1}{2}} -\lambda\boldsymbol W_0^{\frac{1}{2}}\boldsymbol h_{rd}^\dag\boldsymbol h_{rd}^T\boldsymbol W_0^{\frac{1}{2}}\\
-\sum_{k=1}^K\theta_k\left(\frac{1}{\tau}-1\right)\boldsymbol W_0^{\frac{1}{2}}\boldsymbol h_{re,k}^\dag\boldsymbol h_{re,k}^T\boldsymbol W_0^{\frac{1}{2}}
\end{array}\right]_{\rm offd}=\boldsymbol 0.
\end{equation}
By subtracting \eqref{eq:off-diagonal cancellation} from \eqref{eq:Y1 intermediate1}, \(\boldsymbol Y_1\) can be rewritten as
\begin{align}
\boldsymbol Y_1 =& -P_s\tilde{\boldsymbol h}_{sd}^\dag\tilde{\boldsymbol h}_{sd}^T+\lambda\boldsymbol D_{\hat{sd}}-\sum_{k=1}^K\theta_k\left(\frac{1}{\tau}-1\right)\boldsymbol D_{\hat{se},k}
-\left[\begin{array}{r}
\lambda\boldsymbol W_0^{\frac{1}{2}}\boldsymbol h_{rd}^\dag\boldsymbol h_{rd}^T\boldsymbol W_0^{\frac{1}{2}}\\
-\sum_{k=1}^K\theta_k\left(\frac{1}{\tau}-1\right)\boldsymbol W_0^{\frac{1}{2}}\boldsymbol h_{re,k}^\dag\boldsymbol h_{re,k}^T\boldsymbol W_0^{\frac{1}{2}}
\end{array}\right]_{\rm d}\notag\\
&+\left[\boldsymbol W_0^{\frac{1}{2}}\boldsymbol Y_2\boldsymbol W_0^{\frac{1}{2}}\right]_{\rm d}+\sum_{k=1}^K\theta_k P_s\tilde{\boldsymbol h}_{se,k}^\dag\tilde{\boldsymbol h}_{se,k}^T,\label{eq:Y1 intermediate2}
\end{align}
where \([\cdot]_{\rm d}\) denotes a square matrix with only the diagonal remained. Observing that
\begin{align}
\!\!(\boldsymbol D_{\hat{sd}})^{-1}\boldsymbol D_{\hat{se}}&=\left[\boldsymbol W_0^{\frac{1}{2}}\boldsymbol h_{rd}^\dag\boldsymbol h_{rd}^T\boldsymbol W_0^{\frac{1}{2}}\right]_{\rm d}^{-1}\left[\boldsymbol W_0^{\frac{1}{2}}\boldsymbol h_{re,k}^\dag\boldsymbol h_{re,k}^T\boldsymbol W_0^{\frac{1}{2}}\right]_{\rm d}\nonumber\\
&= {\rm diag}\left([|h_{r_ie,k}|^2\left.\middle/\right.|h_{r_id}|^2]_{i=1}^N\right)\notag\\
&\equiv \boldsymbol R_{ed,k},\label{eq:R_{ed,k}}
\end{align}
\(\boldsymbol Y_1^\ast\) can be finally recast as
\begin{align}
\boldsymbol Y_1^\ast=-P_s\tilde{\boldsymbol h}_{sd}^\dag\tilde{\boldsymbol h}_{sd}^T+\boldsymbol\Xi+\sum_{k=1}^K\theta_kP_s\tilde{\boldsymbol h}_{se,k}^\dag\tilde{\boldsymbol h}_{se,k}^T,
\label{eq:Y1 final}
\end{align}
where
\begin{align}\label{eq:Xi}
\boldsymbol\Xi=\left[\boldsymbol W_0^{\frac{1}{2}}\boldsymbol Y_2\boldsymbol W_0^{\frac{1}{2}}\right]_{\rm d}-\left(\left[\boldsymbol W_0^{\frac{1}{2}}\boldsymbol h_{rd}^\dag\boldsymbol h_{rd}^T\boldsymbol W_0^{\frac{1}{2}}\right]_{\rm d}-\boldsymbol D_{\hat{sd}}\right)
\left(\lambda\boldsymbol I-\sum_{k=1}^K\theta_k\left(\frac{1}{\tau}-1\right)\boldsymbol R_{ed,k}\right).
\end{align}

In the following, we show that \(\boldsymbol\Xi+\sum_{k=1}^K\theta_k P_s\tilde{\boldsymbol h}_{se,k}^\dag\tilde{\boldsymbol h}_{se,k}^T\) is a positive definite matrix. Note that since \(\boldsymbol\Xi\) is a diagonal matrix, its definiteness is only determined by the signs of its diagonal entries, for which we commence with the discussion in three difference cases below.

\begin{enumerate}
\item[1)]{\bf Case I}: \(\exists i\) such that \(\lambda-\sum_{k=1}^K\theta_k\left(\frac{1}{\tau}-1\right)[\boldsymbol R_{ed,k}]_{i,i}<0\). Since \(\left[\left[\boldsymbol W_0^{\frac{1}{2}}\boldsymbol h_{rd}^\dag\boldsymbol h_{rd}^T\boldsymbol W_0^{\frac{1}{2}}\right]_{\rm d}-\boldsymbol D_{\hat{sd}}\right]_{i,i}=\)\\ \(\eta\bar\alpha_i P_s\vert h_{sr_i}\vert^2\vert h_{r_id}\vert^2\left(1-\frac{(1-\bar\alpha_i)\sigma_{n_a}^2+\sigma_{n_c}^2}{(1-\bar\alpha_i)(\vert h_{sr_i}\vert^2 P_s+\sigma_{n_a}^2)+\sigma_{n_c}^2}\right)>0\), it follows from \eqref{eq:Xi} that \([\boldsymbol\Xi]_{i,i}>0\) in this case.
\item[2)]{\bf Case II}: \(\exists i\) such that \(\lambda-\sum_{k=1}^K\theta_k\left(\frac{1}{\tau}-1\right)[\boldsymbol R_{ed,k}]_{i,i}>0\). We have \(\left[\left[\boldsymbol W_0^{\frac{1}{2}}\boldsymbol Y_2\boldsymbol W_0^{\frac{1}{2}}\right]_{\rm d}\right]_{i,i}-[\boldsymbol W_0]_{i,i}\vert h_{r_id}\vert^2(\lambda^\ast-\sum_{k=1}^K\theta_k\left(\frac{1}{\tau}-1\right)[\boldsymbol R_{ed,k}]_{i,i})\ge 0\) in accordance with \eqref{eq:Y2 for SPS}, which implies that \([\boldsymbol\Xi]_{i,i}=\left[\left[\boldsymbol W_0^{\frac{1}{2}}\boldsymbol Y_2\boldsymbol W_0^{\frac{1}{2}}\right]_{\rm d}\right]_{i,i}-[\boldsymbol W_0]_{i,i}\vert h_{r_id}\vert^2(\lambda-\sum_{k=1}^K\theta_k\left(\frac{1}{\tau}-1\right)[\boldsymbol R_{ed,k}]_{i,i})+[\boldsymbol D_{\hat{sd}}]_{i,i}(\lambda-\sum_{k=1}^K\theta_k\left(\frac{1}{\tau}-1\right)[\boldsymbol R_{ed,k}]_{i,i})>0\) (c.f.~\eqref{eq:Xi}).
\item[3)]{\bf Case III}: \(\exists i\) such that \(\lambda-\sum_{k=1}^K\theta_k\left(\frac{1}{\tau}-1\right)[\boldsymbol R_{ed,k}]_{i,i}=0\). In this case, it follows that \([\boldsymbol\Xi]_{i,i}= \left[\boldsymbol W_0^{\frac{1}{2}}\boldsymbol Y_2\boldsymbol W_0^{\frac{1}{2}}\right]_{i,i}\)\\ \(\ge 0\). It is noteworthy that the number of \(i\)'s such that \(\lambda-\sum_{k=1}^K\theta_k\left(\frac{1}{\tau}-1\right)[\boldsymbol R_{ed,k}]_{i,i}=0\) cannot exceed one. This can be proved by contradiction as follows. If \(\exists i_1,i_2\), \(i_1\neq i_2\), such that \(\lambda-\sum_{k=1}^K\theta_k\left(\frac{1}{\tau}-1\right)[\boldsymbol R_{ed,k}]_{i_1,i_1}=0\) and \(\lambda-\sum_{k=1}^K\theta_k\left(\frac{1}{\tau}-1\right)[\boldsymbol R_{ed,k}]_{i_2,i_2}=0\), it implies that \(\sum_{k=1}^K\theta_k[\boldsymbol R_{ed,k}]_{i_1,i_1}=\sum_{k=1}^K\theta_k[\boldsymbol R_{ed,k}]_{i_2,i_2}\), which contradicts to the fact that for any two independent continuously distributed random variables, the chance that they are equal is zero.
\end{enumerate}

In summary, \([\boldsymbol\Xi]_{i,i}\ge 0\), \(\forall i\). If \([\boldsymbol\Xi]_{i,i}> 0\), then it is obvious that \(\boldsymbol\Xi+\sum_{k=1}^K\theta_k P_s\tilde{\boldsymbol h}_{se,k}^\dag\tilde{\boldsymbol h}_{se,k}^T\succ\boldsymbol 0\). Next, we show that it still holds true in the case that \(\exists i^\prime\), such that \([\boldsymbol\Xi]_{i^\prime,i^\prime}=0\), \(i^\prime\in\mathcal{N}\), by definition. Define the null-space of \(\boldsymbol\Xi\) by \(\psi=\{\boldsymbol\eta\vert\boldsymbol\eta=\alpha\boldsymbol e_{i^\prime},\alpha\in\mathbb{C}\}\) and multiply \(\boldsymbol\eta^H\) and \(\boldsymbol\eta\), \(\forall\boldsymbol\eta\neq\boldsymbol 0\), on the LHS and RHS of \(\boldsymbol\Xi+\sum_{k=1}^K\theta_k P_s\tilde{\boldsymbol h}_{se,k}^\dag\tilde{\boldsymbol h}_{se,k}^T\), respectively. If \(\boldsymbol\eta\notin\psi\), it is straightforward to obtain \(\boldsymbol\eta^H(\boldsymbol\Xi+\sum_{k=1}^K\theta_k P_s\tilde{\boldsymbol h}_{se,k}^\dag\tilde{\boldsymbol h}_{se,k}^T)\boldsymbol\eta>0\); otherwise, it follows that \(\boldsymbol\eta^H(\boldsymbol\Xi+\sum_{k=1}^K\theta_k P_s\tilde{\boldsymbol h}_{se,k}^\dag\tilde{\boldsymbol h}_{se,k}^T)\boldsymbol\eta=\sum_{k=1}^K\theta_k P_s\alpha^2\vert[\tilde{\boldsymbol h}_{se,k}]_{i^\prime}\vert^2>0\), as \([\tilde{\boldsymbol h}_{se,k}]_{i^\prime}\neq 0\) in probability. This completes the proof. As a result, \(\boldsymbol Y_1\) in \eqref{eq:Y1 final} is shown to always take on a special structure, that is, a full-rank matrix minus a rank-one matrix. Note that this observation plays a key role in proving the rank-one property of \(\hat{\boldsymbol X}_1^\ast\), which is also identified in \cite[Appendix C]{Li2015}.

\begin{figure*}
\begin{align}
\left\|\begin{array}{c}2\sigma_{n_c}{\rm trace}(\boldsymbol U_2\boldsymbol E_i)\\
2\sqrt{\left(1-\frac{\bar z_i}{c_{0,i}}\right)\frac{1}{c_{1,i}}}\\
\left(1-\frac{\bar z_i}{c_{0,i}}-c_{1,i}{\rm trace}(\boldsymbol U_1\boldsymbol E_i)\right)-\left({\rm trace}(\boldsymbol U_2\boldsymbol E_i)+\frac{1}{c_{1,i}}\right)
\end{array}\right\|&\le \left(1-\frac{\bar z_i}{c_{0,i}}-c_{1,i}x_i\right)+\left(y_i+\frac{1}{c_{1,i}}\right)\label{eq:equivalent SOC per-relay jamming constraint}\\
\left\|\begin{array}{c}2\sigma_{n_c}{\rm trace}(\boldsymbol U_2\boldsymbol E_i)\\
2\sqrt{\frac{1}{c_{1,i}}}\\
\left(1-c_{1,i}{\rm trace}(\boldsymbol U_1\boldsymbol E_i)\right)-\left({\rm trace}(\boldsymbol U_2\boldsymbol E_i)+\frac{1}{c_{1,i}}\right)
\end{array}\right\|&\le \left(1-c_{1,i}x_i\right)+\left(y_i+\frac{1}{c_{1,i}}\right)\label{eq:equivalent SOC constraint on rho}
\end{align}
\hrulefill
\end{figure*}

Finally, multiplying both sides of \eqref{eq:Y1 final} by \(\hat{\boldsymbol X}_1^\ast\), as per \eqref{eq:CS wrt hat X1}, we obtain the following equation:
\begin{equation}\label{eq:rank-one structure of hat X1}
 \hat{\boldsymbol X}_1^\ast=P_s\left(\boldsymbol\Xi+\sum_{k=1}^K\theta_k P_s\tilde{\boldsymbol h}_{se,k}^\dag\tilde{\boldsymbol h}_{se,k}^T\right)^{-1}\tilde{\boldsymbol h}_{sd}^\dag\tilde{\boldsymbol h}_{sd}^T\hat{\boldsymbol X}_1^\ast,
\end{equation}
which further implies that \({\rm rank}( \hat{\boldsymbol X}_1^\ast)\le{\rm rank}(\tilde{\boldsymbol h}_{sd}^\dag\tilde{\boldsymbol h}_{sd}^T)=1\). In addition, since the optimality of \(\mathrm{(P1.1\text{-}SDP)}\)  suggests that \(\hat{\boldsymbol X}_1^\ast\neq\boldsymbol 0\), \({\rm rank}( \hat{\boldsymbol X}_1^\ast)=1\) is thus proved.

As \(\hat{\boldsymbol X}_1^\ast\) can be decomposed as \(\hat{\boldsymbol w}_1^\ast\hat{\boldsymbol w}_1^{\ast H}\) by EVD, \eqref{eq:CS wrt hat X1} results in \(\boldsymbol Y_1\hat{\boldsymbol w}_1^\ast=\boldsymbol 0\), which further implies that
\begin{equation}\label{eq:rank-one structure of hat w1}
 \hat{\boldsymbol w}_1^\ast=P_s\left(\boldsymbol\Xi+\sum_{k=1}^K\theta_k P_s\tilde{\boldsymbol h}_{se,k}^\dag\tilde{\boldsymbol h}_{se,k}^T\right)^{-1}\tilde{\boldsymbol h}_{sd}^\dag\tilde{\boldsymbol h}_{sd}^T\hat{\boldsymbol w}_1^\ast.
\end{equation}
Therefore, \eqref{eq:rank-one structure of hat w1} admits a unique solution \(\hat{\boldsymbol w}_1\) up to a scaling factor, which is given by
\begin{equation}\label{eq:hat w1}
\hat{\boldsymbol w}_1=\left(\boldsymbol\Xi+\sum_{k=1}^K\theta_k P_s\tilde{\boldsymbol h}_{se,k}^\dag\tilde{\boldsymbol h}_{se,k}^T\right)^{-1}\tilde{\boldsymbol h}_{sd}^\dag.
\end{equation}
Consequently, we have \(\hat{\boldsymbol w}_1^\ast=\beta\hat{\boldsymbol w}_1\), where \(\beta\in\mathbb{R}_+\). On the other hand, by plugging \(\hat{\boldsymbol w}_1^\ast=\beta\hat{\boldsymbol w}_1\) into the equality constraint of \(\mathrm{(P1.1\text{-}SDP)}\), we have \(\beta=\sqrt{\frac{\tau-\xi^\ast\sigma_{n_d}^2-{\rm trace}(\hat{\boldsymbol S}^\ast\boldsymbol h_{rd}^\dag\boldsymbol h_{rd}^T)}{{\rm trace}(\hat{\boldsymbol w}_1\hat{\boldsymbol w}_1^H\boldsymbol D_{\hat{sd}})}}\), which yields
\begin{equation}\label{eq: hat w1*}
\hat{\boldsymbol w}_1^\ast=\sqrt{\frac{\tau-\xi^\ast\sigma_{n_d}^2-{\rm trace}(\hat{\boldsymbol S}^\ast\boldsymbol h_{rd}^\dag\boldsymbol h_{rd}^T)}{{\rm trace}(\hat{\boldsymbol w}_1\hat{\boldsymbol w}_1^H\boldsymbol D_{\hat{sd}})}}\hat{\boldsymbol w}_1.
\end{equation}

At last, we show $3)$ of Proposition~\ref{prop:rank-one solution with CJ for SPS}. For the case of \(K\ge N\), it is obvious that \({\rm rank}(\hat{\boldsymbol S})^\ast\le N\). For the case of \(K<N\), only a sketch of the proof is provided here due to the length constraint. According to \eqref{eq:Y2 for SPS}, first it is provable that \(\lambda\boldsymbol h_{rd}^\dag\boldsymbol h_{rd}^T+\boldsymbol U\) is a full-rank matrix when \(\mathrm{(P1.1\text{-}SDP)}\) obtains its optimum value; next, observing that \({\rm rank}(\boldsymbol Y_2)\ge N-{\rm rank}(\sum_{k=1}^K\theta_k\boldsymbol h_{re,k}^\dag\boldsymbol h_{re,k}^T)\), it follows that \({\rm rank}(\boldsymbol Y_2)\ge N-K\) as a result of \({\rm rank}(\sum_{k=1}^K\theta_k\boldsymbol h_{re,k}^\dag\boldsymbol h_{re,k}^T)\le K\); then according to \eqref{eq:CS wrt hat S}, \({\rm rank}(\hat{\boldsymbol S}^\ast)\le K\) is thus obtained.

\subsection{Proof of Proposition~\ref{prop:rank-one solution with CJ for DPS}}\label{appendix:proof of rank-one solution with CJ for DPS}
First, use the following lemma to rewrite \(\mathrm{(P2^\prime.1\text{-}SDR)}\).
\begin{lemma}\label{lemma:equivalence of (P2'.1-SDR)}
  Problem \(\mathrm{(P2^\prime.1\text{-}SDR)}\) is equivalent to the following problem:
\begin{align*}
&~\mathrm{(P2^\prime.1\text{-}SDR\text{-}Eqv)}:\\
&\left\{\begin{aligned}
&\max_{\boldsymbol U_1,\boldsymbol U_2,\{x_i\},\{y_i\}}~\eqref{eq:obj for (P2'.1-SDR)} \\
&\mathtt {s.t.}~\eqref{eq:SINR constraint for (P2'.1-SDR)}, \; \forall k, \; \eqref{eq:equivalent SOC per-relay jamming constraint}, \, \eqref{eq:equivalent SOC constraint on rho}, \; \forall i,\\
&{\rm trace}(\boldsymbol U_1\boldsymbol E_i)\le x_i, \, {\rm trace}(\boldsymbol U_2\boldsymbol E_i)\ge y_i,\; \forall i,\\
&{\rm trace}((\boldsymbol U_1-\boldsymbol U_2)\boldsymbol E_i)\le 0, \; \forall i,\\
&\boldsymbol U_1\succeq \boldsymbol 0, \ \boldsymbol U_2\succeq \boldsymbol 0,
\end{aligned}\right.
\end{align*}
where \eqref{eq:equivalent SOC per-relay jamming constraint} and \eqref{eq:equivalent SOC constraint on rho} are given at the top of next page.
\end{lemma}

\begin{IEEEproof}
For convenience of the proof, the optimum value for \(\mathrm{(P2^\prime.1\text{-}SDR)}\) and \(\mathrm{(P2^\prime.1\text{-}SDR\text{-}Eqv)}\) are denoted by \(f_0^\ast\) and \(\tilde f_0^\ast\), respectively. Assuming that \((\boldsymbol U_1^\ast, \boldsymbol U_2^\ast, \{x_i^\ast\}, \{y_i^\ast\})\) is the optimal solution to \(\mathrm{(P2^\prime.1\text{-}SDR)}\), it is easily verified to be feasible for \(\mathrm{(P2^\prime.1\text{-}SDR\text{-}Eqv)}\) as well, which implies that \(f_0^\ast\le\tilde f_0^\ast\). On the other hand, if \(\mathrm{(P2^\prime.1\text{-}SDR\text{-}Eqv)}\) returns an optimal solution of \((\tilde{\boldsymbol U}_1^\ast, \tilde{\boldsymbol U}_2^\ast, \{\tilde x_i^\ast\}, \{\tilde y_i^\ast\})\), by defining \({\rm trace}(\tilde{\boldsymbol U}_1^\ast\boldsymbol E_i)=x_i^{\prime\ast}\) and \({\rm trace}(\tilde{\boldsymbol U}_2^\ast\boldsymbol E_i)=y_i^{\prime\ast}\), \(\forall i\), we show that \((\tilde{\boldsymbol U}_1^\ast, \tilde{\boldsymbol U}_2^\ast, \{x_i^{\prime\ast}\}, \{y_i^{\prime\ast}\})\) is also feasible for \(\mathrm{(P2^\prime.1\text{-}SDR)}\) as follows. As for \eqref{eq:SOC per-relay jamming constraint},
\begin{align}
&\left\|\begin{array}{c}2\sigma_{n_c}y_i^{\prime\ast}\\
2\sqrt{\left(1-\frac{\bar z_i}{c_{0,i}}\right)\tfrac{1}{c_{1,i}}}\\
\left(1-\frac{\bar z_i}{c_{0,i}}-c_{1,i}x_i^{\prime\ast}\right)-\left(y_i^{\prime\ast}+\frac{1}{c_{1,i}}\right)
\end{array}\right\|\notag\\
&\stackrel{(a)}{\le} \left(1-\frac{\bar z_i}{c_{0,i}}-c_{1,i}\tilde x_i^\ast\right)+\left(\tilde y_i^\ast+\frac{1}{c_{1,i}}\right)\notag\\
&\stackrel{(b)}{\le}\left(1-\tfrac{\bar z_i}{c_{0,i}}-c_{1,i}{\rm trace}(\tilde{\boldsymbol U}_1^\ast\boldsymbol E_i)\right)+\left({\rm trace}(\tilde{\boldsymbol U}_2^\ast\boldsymbol E_i)+\tfrac{1}{c_{1,i}}\right)\notag\\
&=\left(1-\frac{\bar z_i}{c_{0,i}}-c_{1,i}x_i^{\prime\ast}\right)+\left(y_i^{\prime\ast}+\frac{1}{c_{1,i}}\right),
\end{align}
where \((a)\) is due to \eqref{eq:equivalent SOC per-relay jamming constraint}, and \((b)\) comes from \({\rm trace}(\tilde{\boldsymbol U}_1^\ast\boldsymbol E_i)\le \tilde x_i^\ast\) and \({\rm trace}(\tilde{\boldsymbol U}_2^\ast\boldsymbol E_i)\ge \tilde y_i^\ast\). Similarly, \((\tilde{\boldsymbol U}_1^\ast, \tilde{\boldsymbol U}_2^\ast, \{x_i^{\prime\ast}\},\)\\ \( \{y_i^{\prime\ast}\})\) can be proved to satisfy \eqref{eq:SOC constraint on rho} as well. In addition, \(x_i^{\prime\ast}-y_i^{\prime\ast}={\rm trace}(\tilde{\boldsymbol U}_1^\ast\boldsymbol E_i)-{\rm trace}(\tilde{\boldsymbol U}_2^\ast\boldsymbol E_i)\le 0\), \(\forall i\), i.e., \eqref{eq:linear constraint on alpha} holds true. These feasibility implies  that \(\tilde f_0^\ast\le f_0^\ast\). By combining the above two facts, we have \(\tilde f_0^\ast=f_0^\ast\) and complete the proof.
\end{IEEEproof}

Then, we apply the Charnes-Cooper transformation again to \(\mathrm{(P2^\prime.1\text{-}SDR\text{-}Eqv)}\), the result of which is denoted by \(\mathrm{(P2^\prime.1\text{-}SDP\text{-}Eqv)}\), to study the property of \(\hat{\boldsymbol U}_1^\ast\) and \(\hat{\boldsymbol U}_2^\ast\). It is noteworthy that the Charnes-Cooper transformed constraint of \eqref{eq:equivalent SOC per-relay jamming constraint} admits the form given by \(\|\boldsymbol x^{(i)}\|\le h(\hat x_i, \hat y_i)\), \(\forall i\),  where \(\boldsymbol x^{(i)}\) is the column vector inside \(\|\cdot\|\) of the LHS of \eqref{eq:equivalent SOC per-relay jamming constraint} and \(h(\hat x_i, \hat y_i)\) indicates the RHS expression.
Since it is easily checked that \(\xi>0\) as a result of feasibility, we have \(\|\boldsymbol x^{(i)}\|>0\Rightarrow h(\hat x_i, \hat y_i)>0\), which implies that
\begin{align}
  \begin{bmatrix}
 h(\hat x_i, \hat y_i)&\boldsymbol x^{(i)^H} \\
 \boldsymbol x^{(i)} & h(\hat x_i, \hat y_i)\boldsymbol I
\end{bmatrix}\succeq\boldsymbol 0 \label{eq:SDP form of the SOC constraint}
\end{align}
according to Schur complement. It thus follows that \eqref{eq:SDP form of the SOC constraint} holds true, \(\forall\boldsymbol x^{(i)}\): \(\|\boldsymbol x^{(i)}\|\le h(\hat x_i, \hat y_i)\). As such, we can show that \eqref{eq:equivalent SOC per-relay jamming constraint} can be recast into a constraint without \(\boldsymbol x\), i.e., not related to \(\hat{\boldsymbol U}_1\), \(\hat{\boldsymbol U}_2\), following the same procedure as \cite[Appendix III]{Chu2015} by exploiting \cite[{\em Lemma 2}]{Eldar2005}. Similarly, the Charnes-Cooper transformed constraint of \eqref{eq:equivalent SOC constraint on rho} can also be rewritten without \(\hat{\boldsymbol U}_1\), \(\hat{\boldsymbol U}_2\). Hence, the partial Lagrangian for \(\mathrm{(P2^\prime.1\text{-}SDP\text{-}Eqv)}\) in terms of \(\hat{\boldsymbol U}_1\) and \(\hat{\boldsymbol U}_2\) can be expressed as (\ref{eq:Lagrangian of (P2'.1-SDP-Eqv)}) (see top of next page),
\begin{figure*}
\begin{multline}\label{eq:Lagrangian of (P2'.1-SDP-Eqv)}
\mathcal{L}(\varphi)=
{\rm trace}\left(\left(\begin{array}{r}
P_s\boldsymbol s_{sd}^\dag\boldsymbol s_{sd}^T-\lambda \sigma_{n_a}^2{\rm diag}(\boldsymbol c_0\circ \|\boldsymbol h_{rd}\|.^2)\\
+\sum_{k=1}^K\theta_k\left(\frac{1}{\tau}-1\right)\sigma_{n_a}^2{\rm diag}(\boldsymbol c_0\circ \|\boldsymbol h_{re,k}\|.^2)\\
-P_s\sum_{k=1}^K\theta_k\boldsymbol s_{se,k}^\dag\boldsymbol s_{se,k}^T
-\boldsymbol\Delta-\boldsymbol\Sigma+\boldsymbol Y_1
\end{array}\right)\hat{\boldsymbol U}_1\right)\\
+{\rm trace}\left(\left(\begin{array}{r}
-\lambda\sigma_{n_c}^2{\rm diag}(\boldsymbol c_0\circ\|\boldsymbol h_{rd}\|.^2)\\
+\sum_{k=1}^K\theta_k\left(\frac{1}{\tau}-1\right)\sigma_{n_c}^2{\rm diag}(\boldsymbol c_0\circ\|\boldsymbol h_{re,k}\|.^2)\\
+\boldsymbol\Pi+\boldsymbol\Sigma+\boldsymbol Y_2
\end{array}\right)\hat{\boldsymbol U}_2\right)
+\lambda\tau,
\end{multline}
\hrulefill
\end{figure*}
where \(\varphi\) denotes a tuple comprising all the associated primal and dual variables: \(\boldsymbol Y_1\), \(\boldsymbol Y_2\), and \(\{\theta_k\}\) are Lagrangian multipliers associated with \(\hat{\boldsymbol U}_1\), \(\hat{\boldsymbol U}_2\), and \eqref{eq:SINR constraint for (P2'.1-SDR)}, \(\forall k\), respectively; \(\lambda\) is the dual variable associated with the only equality constraint; \(\Delta={\rm diag}([\delta_i]_{i=1}^N)\) and \(\Pi={\rm diag}([\pi_i]_{i=1}^N)\) denote those associated with \({\rm trace}(\boldsymbol U_1\boldsymbol E_i)\le x_i\) and \({\rm trace}(\boldsymbol U_2\boldsymbol E_i)\ge y_i\), \(\forall i\), respectively; finally, the diagonal entry of \(\Sigma={\rm diag}([\sigma_i]_{i=1}^N)\) denotes the dual variable associated with \({\rm trace}((\boldsymbol U_1-\boldsymbol U_2)\boldsymbol E_i)\le 0\), \(\forall i\). The KKT conditions related to \eqref{eq:Lagrangian of (P2'.1-SDP-Eqv)} are accordingly given by
\begin{subequations}\label{eq:KKT conditions for (P2'.1-SDP-Eqv)}
\begin{align}
&\boldsymbol Y_1=-P_s\boldsymbol s_{sd}^\dag\boldsymbol s_{sd}^T+\boldsymbol\Xi^\prime+P_s\sum_{k=1}^K\theta_k\boldsymbol s_{se,k}^\dag\boldsymbol s_{se,k}^T \label{eq:Y1 for DPS}\\
&\boldsymbol Y_2=\boldsymbol D-\boldsymbol\Pi-\boldsymbol\Sigma,\label{eq:Y2 for DPS}\\
&\boldsymbol Y_1\hat{\boldsymbol U}_1^\ast=\boldsymbol 0,\label{eq:CS wrt hat U1}\\
&\boldsymbol Y_2\hat{\boldsymbol U}_2^\ast=\boldsymbol 0,\label{eq:CS wrt hat U2}
\end{align}
\end{subequations}
where we have introduced \(\boldsymbol\Xi^\prime= \frac{\sigma_{n_a}^2}{\sigma_{n_c}^2}\boldsymbol D+\boldsymbol\Sigma+\boldsymbol\Delta\), and \(\boldsymbol D=\lambda\sigma_{n_c}^2{\rm diag}(\boldsymbol c_0\circ\|\boldsymbol h_{rd}\|.^2)-\sum_{k=1}^K\theta_k\left(\frac{1}{\tau}-1\right)\sigma_{n_c}^2{\rm diag}(\boldsymbol c_0\circ\|\boldsymbol h_{re,k}\|.^2)\) for notation simplicity. Next, we show that \(\boldsymbol\Xi^\prime+P_s\sum_{k=1}^K\theta_k\boldsymbol s_{se,k}^\dag\boldsymbol s_{se,k}^T\) in \eqref{eq:Y1 for DPS} is a positive definite matrix in the following two cases.
\begin{enumerate}
\item[1)]{\bf Case I}: \(\theta_k=0\), \(\forall k\in\mathcal{K}\). In this case, since \(\lambda>0\) (c.f.~\eqref{eq:Lagrangian of (P2'.1-SDP-Eqv)}) due to the strong duality, it is easily verified that \(\boldsymbol D\succ\boldsymbol 0\) and therefore \( \boldsymbol\Xi^\prime\succ\boldsymbol 0\).
\item[2)]{\bf Case II}: \(\exists k\) such that \(\theta_k\neq0\). According to \eqref{eq:Y2 for DPS}, \(\boldsymbol Y_2\succeq\boldsymbol 0\Rightarrow\boldsymbol D\succeq\boldsymbol\Pi+\boldsymbol\Sigma\), which is a positive semidefinite diagonal matrix. According to the similar argument made in  {\bf Case III} of Appendix~\ref{appendix:proof of rank-one solution with CJ for SPS}, it is thus shown that \(\boldsymbol\Xi^\prime\) has maximum one zero diagonal entry and the positive definiteness of \(\boldsymbol\Xi^\prime+P_s\sum_{k=1}^K\theta_k\boldsymbol s_{se,k}^\dag\boldsymbol s_{se,k}^T\succ\boldsymbol 0\) can be proved by definition without difficulty.
\end{enumerate}

As \(\boldsymbol Y_1\) (c.f.~\eqref{eq:Y1 for DPS}) again complies with the  difference between a positive definite matrix and a rank one matrix, i.e., \(P_s\boldsymbol s_{sd}^\dag\boldsymbol s_{sd}^T\), it turns out that \({\rm rank}(\hat{\boldsymbol U}_1)\le 1\) according to \eqref{eq:CS wrt hat U1}. Then, following the same procedure as that in Appendix~\ref{appendix:proof of rank-one solution with CJ for SPS}, $2)$ of Proposition~\ref{prop:rank-one solution with CJ for DPS} can be proved (details omitted for brevity).

Finally, it is verified that \(\mathrm{(P2^\prime.1\text{-}SDP)}\) is related to \(\hat{\boldsymbol U}_2^\ast\) merely with its diagonal entries, viz \([\hat{\boldsymbol U}_2^\ast]_{i,i}\), \(\forall i\in\mathcal{N}\), (c.f.~\eqref{eq:obj for (P2'.1-SDR)}, \eqref{eq:SINR constraint for (P2'.1-SDR)}). Furthermore, denoting \([[\hat{\boldsymbol U}_2^\ast]_{i,i}^{1/2}]_{i=1}^N\) by \(\hat{\boldsymbol u}_2^\ast\), it is easily checked that the diagonal entries remain the same after we replace \(\hat{\boldsymbol U}_2^\ast\) with \(\hat{\boldsymbol u}_2^\ast\hat{\boldsymbol u}_2^{\ast H}\). Combining the above two facts, we arrive at the conclusion that such modification does not affect its optimality while returning a rank-one \(\hat{\boldsymbol U}_2^\ast\) for \(\mathrm{(P2^\prime.1\text{-}SDP)}\), which completes the proof for $3)$.
\end{appendix}

\end{document}